\documentclass{aastex}
\usepackage{emulateapj5,psfig}
\newcommand \Lya          {\hbox{Ly$\alpha$}}
\newcommand \Zsun          {\hbox{Z$_{\odot}$}}
\newcommand \Msun          {\hbox{M$_{\odot}$}}
\begin{document}
\title{Active Galactic Nuclei Emission-Line Properties Versus Eddington Ratio}
\author{Craig Warner\altaffilmark{1}, Fred Hamann\altaffilmark{1}, 
and Matthias Dietrich\altaffilmark{2}}
\altaffiltext{1}{Department of Astronomy, University of Florida,
211 Bryant Space Science Center, Gainesville, FL 32611-2055,
\\E-mail: warner@astro.ufl.edu, hamann@astro.ufl.edu}
\altaffiltext{2}{Department of Physics and Astronomy, Georgia State
University, Atlanta, Georgia 30303}
\begin{abstract}
We analyze UV spectra for a large sample of 578 Type 1 Active Galactic Nuclei
and derive Eddington ratios, $L/L_{edd}$, from the bolometric luminosities
and emission line widths for each object in the sample.  The sample spans
five orders of magnitude in supermassive black hole (SMBH) mass, seven orders
of magnitude in luminosity, and a redshift range from $0 \leq z \leq 5$.
We include a sample of 26 low-redshift Narrow-Line Seyfert 1s (NLS1s) for
comparative analysis.  The NLS1s have slightly larger than average
$L/L_{edd}$ ratios (and smaller SMBH masses) for their luminosities, but
those $L/L_{edd}$ values are still substantially below the average for
luminous quasars.  A large fraction (27\%) of the objects overall have
$L/L_{edd} > 1$, which might be explained by
non-spherically symmetric accretion.  We find no trend between
$L/L_{edd}$ and either redshift or SMBH mass.  Composite spectra sorted by
$L/L_{edd}$ show an unusual emission-line behavior: nearly constant peak
heights and decreasing FWHMs with increasing $L/L_{edd}$.  This is in marked
contrast to the emission-line behaviors with luminosity, SMBH mass, and
FWHM(\ion{C}{4}), which clearly show trends analogous to the Baldwin Effect:
decreasing line peaks and equivalent widths with increasing luminosity, SMBH
mass, and FWHM.  The origins of the unusual behavior with $L/L_{edd}$ are not
understood, but one implication is that metallicity estimates
based on emission line ratios involving nitrogen show no trend with
$L/L_{edd}$ in the composite spectra created from different ranges in
$L/L_{edd}$.  The NLS1 composite, however, shows a slightly high metallicity
for its SMBH mass and luminosity.  Our earlier work suggests that host galaxy
mass, correlated with SMBH mass and AGN luminosity, is the fundamental
parameter affecting BLR metallicities.  Some secondary effect, not related
to $L/L_{edd}$, must be enhancing the metallicities in NLS1s.
\end{abstract}

\section{Introduction}
	The central engines of quasars, and more generally, active galactic
nuclei (AGNs) are believed to be powered by supermassive black holes (SMBHs).
Two of the fundamental properties of AGNs are the SMBH mass and the accretion
rate of material onto the SMBH.  Several indirect methods have been devised
to estimate SMBH masses.  One set of these methods assumes that the broad
emission-line region (BLR) is in gravitational equilibrium with the central
source, so that the SMBH mass can be estimated by applying the virial
theorem, $M_{\rm SMBH} = rv^2/G$, to the measured line widths (Peterson 1993;
Peterson 1997; Wandel et al. 1999; Kaspi et al. 2000; McLure \& Dunlop 2001;
Vestergaard 2002).  In reverberation
mapping studies, $R_{\rm BLR}$, the radial distance between the central source
and the BLR can be estimated from the lag time between continuum variations
and the emission-line response (Peterson 1993; Peterson 1997; Wandel et
al. 1999; Kaspi et al.  2000).  These reverberation mapping studies have
demonstrated an observed relation of $R_{\rm BLR}~\propto~\lambda~
L_{\lambda}$(5100~\AA)$^{0.7}$ that can be used to estimate $R_{BLR}$ for
AGNs over a wide range of redshifts (Kaspi et al. 2000; McLure \& Dunlop 2001; 
Vestergaard 2002; Corbett et al. 2003; Warner et al. 2003).  Netzer (2003) 
has argued that the slope is not known to an accuracy better than about 0.15.

	A second set of methods is based on the tight correlation between
the masses of SMBHs and the velocity dispersions, $\sigma$, of their host
galaxy spheroidal components (Ferrarese \& Merritt 2000; Gebhardt et al. 2000;
Merritt \& Ferrarese 2001; Tremaine et al. 2002).  However, stellar velocity
dispersions are not easy to measure for AGN hosts, especially at high
redshifts.  Because of this, methods have been devised using proxies of the
velocity dispersion, such as the width of the narrow emission line
[\ion{O}{3}] $\lambda 5007$ (Nelson 2000; Boroson 2003; Shields et al. 2003)
or the bulge
luminosity, $L_{bulge}$ (Magorrian et al. 1998; Laor 1998; Wandel 1999).
Early studies showed a large scatter, as much as two orders of magnitude
between SMBH mass and $L_{bulge}$ (Ferrarese \& Merritt 2000).  However, more
recent studies that carefully model the bulge light profiles of disk galaxies
and thus obtain more accurate values of $L_{bulge}$ show less scatter in
$M_{\rm SMBH}-L_{\rm bulge}$, similar to that in the $M_{\rm SMBH}-\sigma$
relationship (McLure \& Dunlop 2002; Erwin et al. 2002; Bettoni et al. 2003).
Recently, SMBH mass has also been shown to correlate strongly with
the global structure of bulges and ellipticals, such that more centrally
concentrated bulges have more massive SMBHs.  This relationship is as strong
as the $M_{\rm SMBH}-\sigma$ relationship with comparable scatter (Graham
et al. 2001; Erwin et al. 2002).

	Once the SMBH mass has been estimated, the Eddington luminosity can
be calculated as $L_{edd} = 1.26 \times 10^{38} M_{\rm SMBH}$ (\Msun) ergs
s$^{-1}$ (e.g., Rees 1984; Peterson 1997).  Eddington luminosity is the
limit in which the inward gravitational force acting on the gas exactly
balances the outward radiation force induced by electron scattering.  It can
be thought of as the maximum possible
luminosity for an object of mass $M_{\rm SMBH}$ that is powered by {\it 
spherical} accretion (Peterson 1997).  The Eddington luminosity can be exceeded
if accretion is not spherically symmetric (see \S5, also Osterbrock 1989;
Begelman 2002; Collin et al. 2002; Wang 2003). AGN luminosities should be
directly proportional to the accretion rate, $L \propto \dot{M}_{acc}$, and
therefore the ratio, $L/L_{edd} \propto \dot{M}_{acc}/M$, is an indirect
measure of the accretion rate relative to the critical Eddington value.

	Narrow Line Seyfert 1s (NLS1s) are a subclass of Seyfert 1s that
exhibit distinct
and unusual properties: very narrow broad emission lines ($H \beta$ FWHM $<
2000$ km s$^{-1}$) with [\ion{O}{3}] $\lambda 5007$ / $H \beta$ ratios of less
than 3 (to exclude Seyfert 2s), strong \ion{Fe}{2} emission, and unusually
strong big blue bumps
(Osterbrock \& Pogge 1985; Kuraszkiewicz et al. 2000; Constantin \& Shields
2003).  NLS1s also land at one extreme end of the Boroson \& Green (1992)
Principal Component 1 (PC1).  It has been suggested that PC1 is strongly
correlated with $L/L_{edd}$ (Boroson \& Green 1992; Boroson 2002; Shemmer \&
Netzer 2002; Constantin \& Shields 2003).  Several studies have suggested
that NLS1s have low SMBH masses for their luminosities, and thus very
high Eddington ratios, near 1 (Mathur 2000; Kuraszkiewicz et al. 2000;
Shemmer \& Netzer 2002; Shemmer et al. 2003).  It has also been suggested that
NLS1s have unusually high metallicities for their luminosities (see \S4.4 and
Figure 11 below, Mathur 2000; Shemmer \& Netzer 2002; Shemmer et al. 2003).
Shemmer \& Netzer (2002) find that NLS1s depart from the nominal 
relationship between metallicity and luminosity in AGNs (Hamann \& Ferland
1999; Dietrich et al. 2003, in prep), with some NLS1s indicating metallicities
as high as those measured in high-luminosity, high-redshift quasars.  Because
of their high metallicities and high Eddington ratios, Mathur (2000) proposed
that NLS1s are analogs of high-redshift ($z \gtrsim 4$) quasars, in that they
may both be in an early evolutionary phase, residing in young host galaxies. 

%	We can estimate the metallicity of the gas in the BLR by analyzing
%prominent emission line ratios.  Ratios involving nitrogen (N) lines are
%especially valuable in determining metallicity, $Z$, because of the
%expected ``secondary" N production via the CNO cycle of nucleosynthesis
%in stars, in which N is produced from existing carbon and oxygen (Shields
%1976; Hamann \& Ferland 1992, 1993, 1999; Ferland et al.  1996; Hamann et al.
%2002).  When this secondary N production dominates, the nitrogen abundance
%should scale as N/H $\propto Z^{2}$ or N/O $\propto$ O/H $\propto Z$ (Tinsley
%1980), providing a sensitive metallicity diagnostic even when direct measures
%of $Z = $O/H are not available.  Observations of \ion{H}{2} regions indicate
%that secondary nitrogen production and N/O $\propto$ O/H scaling dominate for
%$Z \gtrsim 0.2 - 0.3$\ \Zsun (van Zee et al. 1998; Pettini et al. 2002).  BLR
%metallicities should be representative of the gas in the central regions of
%galaxies if the BLR was enriched by stars in those environments.  See also 
%Hamann \& Ferland 1999, Hamann et al. 2002, and Warner et al. 2003 for further
%discussion. 

	We have collected a large sample of 578 spectra of ``Type 1" AGNs
(quasars and Seyfert galaxies with broad emission lines) that span the
rest-frame UV wavelengths needed for this study  (Dietrich et al. 2002).  We
compute composite spectra from different ranges in the Eddington ratio,
$L/L_{edd}$.  We include a composite spectrum produced from a subsample of 26
NLS1s for comparative analysis.  We present measurements of the emission lines
in these composite spectra and investigate their relationship to $L/L_{edd}$. 

\section{Eddington Ratio Determinations}
	We first estimate SMBH masses by applying the virial theorem,
$M_{\rm SMBH} = rv^2/G$, to the line-emitting gas (for more details, see
also Kaspi et al. 2000; Peterson \& Wandel 2000; McLure \& Dunlop 2001;
Vestergaard 2002; Corbett et al. 2003; Warner et al. 2003).
%In this equation, $v$ is the velocity dispersion and $r$ is the
%radial distance away from the SMBH.  The relationship between $v$ and
%the FWHM of the broad emission line profile depends on the kinematics
%and geometry of the BLR, but can be expressed as $v = k{\rm FWHM}$, where
%$k$ is a factor on the order of unity (McLure \& Dunlop 2001; Vestergaard
%2002).  We make the assumption that the gas in the BLR is gravitationally
%bound and BLR velocities are random ($k = \sqrt{3}/2$) because it is the
%most basic and allows for comparison with most other work (Kaspi et al. 2000;
%Vestergaard 2002; Warner et al. 2003).  Any offset in mass due to different
%kinematic or geometric assumptions would only affect the absolute SMBH masses
%and Eddington ratios and not any trends derived between these and other
%quantities.
Kaspi et al. (2000) express the SMBH mass as
\begin{equation}
M_{SMBH} = 1.5 \times 10^{5} \Msun \left( {{R_{{\rm BLR}}} \over {{\rm lt-days}}} \right) \left( {{\rm FWHM}} \over {{\rm 10^{3}\ km\ s^{-1}}} \right) ^{2} \
\end{equation}
$R_{\rm BLR}$ is the radial distance between the BLR and the central source,
and FWHM applies to the broad emission line profile.  We estimate $R_{\rm BLR}$
based on the observed relation between $R_{\rm BLR}$ for a particular line
and the continuum luminosity (Wandel et al. 1999; Kaspi et al. 2000;
Vestergaard 2002; Corbett et al. 2003).  A particular line must be specified
because reverberation studies have shown that the BLR is radially stratified,
such that higher ionization lines tend to form closer to the central engine
than lower ionization lines (Peterson 1993).

	We select the \ion{C}{4} $\lambda 1549$ emission line instead of
$H \beta$ to estimate SMBH masses because it is more readily observed
across the entire redshift range from $z \sim 0$ to $z \sim 5$.
%It has been suggested that \ion{Mg}{2} $\lambda 2798$ may be a better
%indicator of SMBH mass than \ion{C}{4} (McLure \& Jarvis 2002).  However,
%\ion{C}{4} is much less prone to significant blending with other emission
%lines and is observable over a wider range of redshifts.
%Finally, we find that there is approximately a
We find that there is approximately a
1:1 correlation between the SMBH mass obtained from \ion{C}{4} and that
obtained from $H \beta$.  There can be significant deviations from this
for individual objects, but the relation holds well for averages of many
objects and for measurements based on composite spectra (see Vestergaard
2002; Warner et al. 2003 for further discussion).

	Reverberation studies indicate that the radius of the BLR for
\ion{C}{4} is about half that of $H \beta$ (Stirpe et al. 1994; Korista et al.
1995; Peterson 1997; Peterson \& Wandel 1999).  We therefore modify the
equation given by Kaspi et al. (2000) to obtain
%to account for this, and we use a powerlaw of the form
%$F_{\nu} \propto \nu^{\alpha}$ with $\alpha = -0.4$ to estimate a conversion
%between the continuum luminosity at 5100 \AA\ and 1450 \AA.  We select
%$\alpha = -0.4$ based on average quasar spectra from Brotherton et al. (2001),
%Vanden Berk et al. (2001), and Dietrich et al. (2002).  This yields
\begin{equation}
R_{\rm BLR}({\rm C IV}) = 9.7 \left[ {{\lambda L_{\lambda}(1450 {\rm \AA})} \over {10^{44} \ {\rm ergs \ s^{-1}}}} \right] ^{0.7} \ {\rm lt-days}
\end{equation}
See Warner et al. (2003) for more details.  From Equations (1) and (2), we
derive 
\begin{equation}
M_{SMBH} = 1.4 \times 10^{6} \Msun \left( {{{\rm FWHM(C IV)}} \over {{\rm 10^{3}\ km\ s^{-1}}}} \right) ^{2} \left( {{\lambda L_{\lambda}(1450 {\rm \AA})} \over {{\rm
10^{44} \ {\rm ergs \ s^{-1}}}}} \right) ^{0.7} \
\end{equation}
Vestergaard (2002) calibrated mass derivations based on \ion{C}{4} against 
estimates using FWHM($H \beta$) and direct measurements of $R_{\rm BLR}(H
\beta)$ from reverberation mapping.  Her technique yields essentially the
same mass relationship (within 10\%) as Equation 3, which helps to confirm
the factor of 2 scaling adopted here between $R_{\rm BLR}$(\ion{C}{4}) and
$R_{\rm BLR}(H \beta)$.   Vestergaard (2002) finds that SMBH masses
estimated by applying this equation to single-epoch spectra of individual
objects have a 1$\sigma$ uncertainty of
a factor of three when compared to studies that use $H \beta$ and a direct,
reverberation measure of the BLR radius.  Our composite spectra average over 
variabilities and object-to-object scatter, which should significantly reduce
the uncertainties.  See also Krolik (2001), Netzer (2003), Corbett et al.
(2003), and Vestergaard (2004) for further discussion of the uncertainties.

	We next estimate bolometric luminosities, $L$, based on an
integration over a typical quasar continuum shape.
We use the cosmological parameters H$_{0}$ = 65 km s$^{-1}$
Mpc$^{-1}$, $\Omega_{M}$ = 0.3, and $\Omega_{\Lambda}$ = 0 (Carroll, Press,
\& Turner 1992) throughout this paper\footnote {The use of this cosmology was
motivated for comparison with earlier studies in the pre-WMAP era.
Furthermore, the difference between our set of cosmological parameters
and those suggested by WMAP results in a difference of less than 20\% in
luminosity for a wide redshift range of $0<z<4$.}.  We assume a segmented
powerlaw of the form $F_{\nu} \propto \nu^{\alpha}$ to approximate the
continuum shape, with $\alpha = -0.9$ from 0.1 \AA\ to 10 \AA, $\alpha =
-1.6$ from 10 \AA\ to 1000 \AA, and $\alpha = -0.4$ from 1000 \AA\ to
100,000 \AA\ (Zheng et al. 1997; Laor et al. 1997; Brotherton et al. 2001;
Vanden Berk et al. 2001; Dietrich et al. 2002).
It is now well established that the mean UV--IR slope is a function of
luminosity (see Dietrich et al. 2002).  For the luminosity range spanned
by our sample, the average powerlaw index between 1000 \AA\ and 100,000 \AA\
ranges from $\sim$ -0.2 to $\sim$ -0.6 (M. Dietrich, private communication).
This range could cause a scatter of $\sim$ 20\% in our estimates of $L$. 

	Integrating this nominal spectrum over all wavelengths implies
bolometric corrections
of 4.36 and 9.27 for $\lambda L_{\lambda}(1450 {\rm \AA})$ and $\lambda
L_{\lambda}(5100 {\rm \AA})$, respectively.  These corrections are slightly
lower than the correction of 11.8 to $\lambda L_{\lambda}(5100 {\rm \AA})$
derived by Elvis et al. (1994), but in good agreement with more recent
derivations (Kaspi et al. 2000; Vestergaard 2004).  We use this bolometric
correction and SMBH masses from Equation (3) to obtain Eddington ratios:
\begin{equation}
{{L} \over {L_{edd}}} = 1.6 \left( {{{\rm FWHM(C IV)}} \over {{\rm 10^{3}\ km\ s^{-1}}}} \right) ^{-2} \left( {{L} \over {{\rm 10^{44}\ ergs \ s^{-1}}}} \right) ^{0.3}
\end{equation} 

\section{Data \& Analysis}

	Our sample is comprised of 578 Type 1 (broad-line) AGN spectra
with rest-frame UV wavelength coverage that encompasses the range $950
\lesssim \lambda \lesssim 2050$ \AA.  The spectra were obtained by several
groups using various ground-based instruments as well as the {\it Hubble Space
Telescope (HST)} and the {\it International Ultraviolet Explorer (IUE)} (see
Dietrich et al. 2002 and 2004, in prep for more details).
The sample spans a redshift range from $0 \lesssim z \lesssim 5$, seven orders
of magnitude in luminosity, and five orders of magnitude in SMBH mass.  One
unique aspect of this sample is that it contains new observations of faint
quasars at redshift $z > 2.5$ (e.g., Steidel et al. 2002, Dietrich et al.
2002).  Thus we can avoid to some degree the bias toward higher luminosities
at higher redshifts inherent in magnitude-limited samples.  The sample spans
at least three orders of magnitude in luminosity at all redshifts (see Fig. 1
in Dietrich et al. 2002).  We determined the radio loudness for the quasars
using the radio flux densities given in V\'{e}ron-Cetty \& V\'{e}ron (2001).
We used the definition of radio loudness given by Kellermann et al. (1989).
Our classifications of radio-loud quasars are consistent with classifications
available in the literature (Wills et al. 1995; Bischof \& Becker 1997;
Wilkes et al. 1999; Stern et al. 2000).

	We use an automated program to estimate the FWHM of \ion{C}{4} in
each spectrum (see Warner et al. 2003 for details).  Comparisons
between the FWHMs estimated by the program and those measured manually
indicate an error of $\lesssim 10$\% in the automated results.  Lines
containing significant absorption are flagged by the program and their
FWHMs are estimated manually (by interpolating across the absorption
feature).  We use the FWHM of
\ion{C}{4} and the continuum luminosity, $\lambda L_{\lambda}(1450 {\rm \AA})$
to estimate the central SMBH mass and $L/L_{edd}$ for each quasar based on
the equations given in \S2.

	We then sort the quasars by $L/L_{edd}$ into seven bins (see Figure
1): $L/L_{edd} < 0.25$, $0.25 \leq L/L_{edd} < 0.33$, $0.33 \leq L/L_{edd} <
0.50$, $0.50 \leq L/L_{edd} < 0.67$, $0.67 \leq L/L_{edd} < 1.00$, $1.00 \leq
L/L_{edd} < 2.00$, and $L/L_{edd} \geq 2.00$, and compute seven composite
spectra.  Each composite spectrum is the average of all the quasar spectra in
a bin.  Table 1 lists various parameters for the composites, including the
mean values of $M_{\rm SMBH}$, $L$, FWHM(\ion{C}{4}), $L/L_{edd}$, and the
redshift, $z$, as measured from the individual objects contributing to each
composite.  Also listed are the numbers of objects contributing at the
wavelength of the \ion{C}{4} emission line.  The spectral slopes, $\alpha$,
are measured from each composite spectrum and constrained by the flux in
20 \AA\ wide windows centered at 1450 \AA\ and 1990 \AA.

	Calculating composite spectra
significantly improves the signal-to-noise ratio and averages over
object-to-object variations.  Since narrow absorption features may affect
the emission line profiles in composite spectra, we developed a method to
detect strong narrow absorption features.  The contaminated spectral region
of the individual spectrum is then excluded from the calculation of the
composite spectrum.  For more details about creating composite spectra,
see Brotherton et al. (2001), Vanden Berk et al. (2001), Dietrich et al.
(2002), and Warner et al. (2003)

        For comparison with the $L/L_{edd}$ composites, we also create
composite spectra for different ranges in SMBH mass ($10^{6}-10^{7}$ \Msun,
$10^{7}-10^{8}$ \Msun, $10^{8}-10^{9}$ \Msun,  $10^{9}-10^{10}$ \Msun, and
$\geq10^{10}$ \Msun; see also Warner et al.  2003), $L$
($10^{44}-10^{45}$ ergs s$^{-1}$, $10^{45}-10^{46}$ ergs s$^{-1}$,
$10^{46}-10^{47}$ ergs s$^{-1}$, $10^{47}-10^{48}$ ergs s$^{-1}$,
and $10^{48}-10^{49}$ ergs s$^{-1}$), and FWHM(\ion{C}{4}) ($<2000$ km s$^{-1}$,
$2000-4000$ km s$^{-1}$, $4000-6000$ km s$^{-1}$, $6000-8000$ km s$^{-1}$,
and $\geq8000$ km s$^{-1}$).  See Table 1 for additional information.

	We also create a composite spectrum of 26 NLS1s that were classified
by others according to the criteria described in \S1 (Kuraszkiewicz et al.
2000; Wang \& Lu 2001; Constantin \& Shields 2003).  This subsample is drawn
from the same overall sample used to create the other composites described
above.  Eighteen of these objects have data at \ion{C}{4} and they
span a range in $L/L_{edd}$ from $\sim$ 0.1 to 2.

	We correct each composite spectrum for strong iron emission lines
using the empirical Fe emission template that was extracted from I\,Zw\,1
by Vestergaard \& Wilkes (2001), which they very kindly provided for this
study (see Dietrich et al. 2002 and Warner et al. 2003 for more details).
The \ion{Fe}{2} contribution is generally small at wavelengths $\lesssim$
2000 \AA, but the correction for this emission improves the measurements
of weak lines such as \ion{N}{3}] $\lambda 1750$ and \ion{He}{2} $\lambda
1640$.  We fit the continuum of each Fe-subtracted spectrum with a
powerlaw of the form $F_{\nu}$~$\propto$~$\nu^{\alpha}$.
Figure 2 shows the final \ion{Fe}{2}-subtracted composite spectra
normalized by the continuum fits.

	To measure the broad emission lines, we use a spectral fitting routine
developed in the IDL language, that employs $\chi^{2}$ minimization.  We
fit each line with one or more Gaussian profiles, with the goal of simply
measuring the total line strengths free of blends.  When necessary, we use
the profile of strong unblended lines, such as \ion{C}{4}, to constrain the
fits to weaker or more blended lines (see Warner et al. 2003 for details
of our fitting procedure).  Figure 3 shows an example of our fits.

	The continuum location is the primary uncertainty in our flux
measurements.  We estimate the 1$\sigma$ standard deviation of our
measurements of the fluxes of \Lya\ and \ion{C}{4} to be $\lesssim$10\%
based on repeated estimates with the continuum drawn at different levels.
We estimate the uncertainty in weaker lines by the same method to be
$\sim$10--20\%.  There are also secondary uncertainties due to line blending,
which can be important for some of the weak lines and for \ion{N}{5} in
the wing of \Lya.

\section{Results \& Comparisons}

\subsection{$L/L_{edd}$ and Super-Eddington Accretion}
	Figure 4 shows the distribution of $L/L_{edd}$ for the entire sample.
A large fraction (27\%) of the objects in our sample have $L/L_{edd} > 1$
(see also Fig. 1).  This result is not precise because our mass estimates
for individual objects have factor of $\sim$3 uncertainties (\S2), which are
comparable to the width of the distribution in Figure 4.  Nonetheless, it
is interesting that the sample mean is close to the Eddington limit, with
$<L/L_{edd}>$ $\sim 0.9$ (Fig. 4).  Also note that trends with $L/L_{edd}$ that
we discuss below are more reliable
than the individual measurements because they rely on relative $L/L_{edd}$.

	Consistent with previous studies (\S1), we find that the NLS1s have
generally high Eddington ratios for their luminosities, including several
objects with $L/L_{edd} \gtrsim 1$.  However, the NLS1s do not have the
highest Eddington ratios in our sample.  Quasars with high luminosities and
narrow \ion{C}{4} emission lines often have $L/L_{edd} > 2$.  The most extreme
of these objects, such as BR2248-1242 (see Warner et al. 2002), can have
derived Eddington ratios approaching 10 (see Figure 1).

\subsection{Correlations with $L/L_{edd}$}
        Figure 5 shows the distributions in redshift, bolometric luminosity,
FWHM(\ion{C}{4}), and SMBH mass as a function of $L/L_{edd}$ for the entire
sample.  $L/L_{edd}$ correlates positively with $L$ and negatively
with FWHM(\ion{C}{4}), but these correlations may be attributed largely to our
derivation of $L/L_{edd}$ from these quantities.  In fact, the slopes in these
correlations are matched well by the parameter relationships in Equation 4.  In
agreement with Woo \& Urry (2002), we find no trend between $L/L_{edd}$ and
either redshift or SMBH mass.  We find that the weak trend in Figure 5 
between $L/L_{edd}$ and redshift is due to i) a trend for larger $L/L_{edd}$
with increasing $L$, and ii) a bias for more high $L$ objects at higher
redshifts in our sample.  Sub-samples spanning narrow ranges in luminosity
show that there is no trend between $L/L_{edd}$ and redshift once these biases
are removed (see Figure 6).  In all four panels of Figure 5, there are no
clear differences between radio loud and radio quiet objects. 

	Table 1 shows that there is no apparent trend between Eddington
ratio and the slope of the UV
continuum.  The NLS1 composite, though, exhibits a steeper (softer) UV
spectrum than the $L/L_{edd}$ composites (see Table 1).  This is consistent
with findings that NLS1s in general have redder spectra than typical Type 1
AGNs (e.g., Crenshaw et al. 2002; Constantin \& Shields 2003).

        Table 2 lists for each $L/L_{edd}$ and NLS1 composite spectrum the
line fluxes relative to \Lya, the rest-frame equivalent widths (REWs) as
measured above the fitted continuum, and the FWHMs.
Figure 7 plots the REWs of selected emission
lines as a function of $L/L_{edd}$.  The NLS1 composite spectrum is displayed
(plotted as a triangle) for comparative purposes.
The dotted lines are linear fits to the $L/L_{edd}$ composite data (excluding
the NLS1 composite).  Interestingly, while most emission
lines decrease in REW with increasing $L/L_{edd}$, \Lya\ and \ion{O}{6}
exhibit a positive trend between REW and $L/L_{edd}$, and \ion{N}{3}] shows
no trend at all.  \ion{O}{6} has larger measurement errors than most other
emission lines, so it is unclear whether this positive trend between
\ion{O}{6} REW and $L/L_{edd}$ is real or not.

        The REWs of emission lines in the NLS1 composite (represented by a
triangle) generally do not match the trend fit to the $L/L_{edd}$ composites
in Figure 7.  The NLS1 composite is above the fitted trend for some emission
lines and below the fitted trend for others, regardless of the slopes of the
trends.
%Neither the origins of the unusual emission-line behavior with
%$L/L_{edd}$ nor the irregular behavoir of the NLS1 composite is understood.

\subsection{Surprising Emission-Line Behaviors}
	The composite spectra sorted by $L/L_{edd}$ show a surprising
emission-line behavior: nearly constant peak heights and decreasing FWHMs with
increasing $L/L_{edd}$ (see Figure 2).  This is in marked contrast to the
emission-line behaviors in composite spectra sorted by luminosity, SMBH
mass, and FWHM(\ion{C}{4}), which clearly show trends analagous to the Baldwin
Effect: decreasing line peaks and equivalent widths with increasing luminosity,
SMBH mass, and FWHM.  Figure 8 compares the emission-line behaviors in these
different composites (see also Wills et al. 1993; Croom et al. 2002;
Dietrich et al.  2002; Warner et al. 2003).
In particular, the composite spectra created from different ranges in
FWHM(\ion{C}{4}) clearly show a trend analagous to the Baldwin Effect despite
spanning a range of less than half an order of magnitude in average
luminosity (Table 1).  In contrast, the $L/L_{edd}$ composites span a wide
range in FWHM(\ion{C}{4}) and an order of magnitude in average luminosity but
do not show any behavior similar to the Baldwin Effect.  This suggests that
the Baldwin Effect may actually be related to SMBH mass (which correlates
positively with both $L$ and FWHM), since the $L/L_{edd}$ composites have
nearly constant $M_{\rm SMBH}$.
%even exhibit this trend for decreasing peak heights
%and equivalent widths with increasing luminosity, despite spanning a much
%narrower range in average luminosity than the $L/L_{edd}$ composites (see
%Tables 1 \& 2).

	To illustrate this point further, Figure 9 compares $L/L_{edd}$
composite spectra created for a narrow range in SMBH mass ($10^8 \Msun <
M_{\rm SMBH} < 10^9 \Msun$) and a narrow luminosity range ($10^{47}$ ergs/s $<
L < 10^{48}$ ergs/s).  Both sets of spectra are shown prior to
normalization to the continuum.  The emission line behavior
described above (e.g. constant peak heights, etc.) is clearly evident in
the $L/L_{edd}$ composites at nearly constant $M_{\rm SMBH}$, but {\it not}
in the composites with nearly constant $L$.  The composites at nearly constant
$L$ show a trend for decreasing peak heights and equivalent widths with
increasing $M_{\rm SMBH}$.

	In composite spectra sorted by luminosity,
SMBH mass, and FWHM(\ion{C}{4}), the Baldwin Effect is not seen in \ion{N}{5}
(see Figure 8, also Dietrich et al. 2002; Warner et al. 2003), leading to
a higher \ion{N}{5} / \ion{C}{4} ratio in objects with higher luminosities,
SMBH masses, and FWHMs(\ion{C}{4}).  However, in the $L/L_{edd}$ composites
created from our entire sample, \ion{N}{5} clearly decreases in REW as
$L/L_{edd}$ increases, yielding a nearly constant \ion{N}{5} / \ion{C}{4}
ratio across the full range of $L/L_{edd}$ (see Figure 10).  Furthermore,
the $L/L_{edd}$ composites created from a narrow range in $M_{\rm SMBH}$
exhibit this behavior of nearly constant \ion{N}{5} / \ion{C}{4}, while
the composites created from a narrow range in $L$ do not (see Fig. 9).  The
composites at nearly constant $L$ show a trend for increasing \ion{N}{5} /
\ion{C}{4} toward lower $L/L_{edd}$ (higher SMBH masses).  The differences
in emission-line behaviors at nearly constant $M_{\rm SMBH}$ and at nearly
constant $L$ have implications for the origin of the Baldwin Effect (see
Warner, Hamann, \& Dietrich 2004, in prep).  Throughout the rest of this
manuscript (including figures and tables), ``$L/L_{edd}$ composites" refer
to the composite spectra created from our entire sample and sorted by
$L/L_{edd}$.

\subsection{Metallicities}
	We compare emission line flux ratios to plots of metallicity vs.
line ratio based on theoretical models (see Figure 5 in Hamann et al. 2002
and Figure 3 in Warner et al. 2002).
Ratios involving nitrogen lines are especially valuable in estimating
metallicity, $Z$, due to the expected ``secondary" N production via the
CNO cycle of nucleosynthesis in stars (Shields 1976; Hamann \& Ferland 1992,
1993, 1999; Ferland et al. 1996; Hamann et al. 2002).  In the CNO cycle, N is
produced from existing carbon and oxygen and thus the nitrogen abundance
scales as N/H $\propto Z^2$ and N/O $\propto$ O/H $\propto Z$ (Tinsley 1980),
providing a sensitive metallicity diagnostic even when direct measurements of
$Z$ are not available (for more discussion, see Wheeler et al. 1989; Hamann \&
Ferland 1999; Henry et al. 2000; Hamann et al. 2003; Pilyugin 2003; Pilyugin
et al. 2003).  We prefer to base our metallicity estimates on the
calculations in Hamann et al.
(2002) that use a segmented powerlaw for the photoionizing continuum shape
because this shape is a good approximation to the average observed continuum
in quasars (Zheng et al. 1997; Laor et al. 1997) and it yields intermediate
results for line ratios that are sensitive to the continuum shape, such
as \ion{N}{5}/\ion{He}{2}.

	Figure 10 shows metallicities inferred from several line ratios as
a function of $L/L_{edd}$.  The uncertainties shown in Figure 10 derive solely 
from the 1$\sigma$ measurement uncertainties discussed in \S3 and do not
include the theoretical uncertainties in the technique we use to derive
metallicities from the line ratios. Our best estimate of the overall
metallicity from each spectrum (labeled as Average in Figure 10) is obtained
by averaging the results of the line ratios that we believe are most
accurately measured and most reliable from a theoretical viewpoint.
Specifically, we average the metallicities derived from
\ion{N}{3}]/\ion{C}{3}], \ion{N}{3}]/\ion{O}{3}], and \ion{N}{5}/\ion{C}{4}
(or when available, \ion{N}{5}/(\ion{C}{4}+\ion{O}{6})).  See Hamann
et al. (2002) and Warner et al. (2003) for further discussion.

	All of the line ratios involving \ion{N}{3}] and \ion{N}{5} show
N/O and N/C ratios that are solar or greater.  This implies a metallicity
of $\gtrsim$ 1 \Zsun\ if N is mostly secondary.  One implication of the
unusual emission-line behavior with $L/L_{edd}$ (discussed in \S4.3) is
that the derived metallicities on average show no trend with $L/L_{edd}$.
The only line ratio to exhibit a strong trend with $L/L_{edd}$ is
\ion{N}{5}/\ion{O}{6}.
%However, because of the large measurement errors in \ion{O}{6}, it is
%not clear whether this apparent trend is real or not.

	Figure 11 shows our best estimates of the overall metallicities
(derived from several line ratios, as described above for the ``Average"
in Figure 10) from the composite spectra sorted by $L/L_{edd}$, SMBH mass, and
luminosity.  The NLS1 composite spectrum is plotted in all three panels for
comparison.  The NLS1s exhibit a metallicity that is slightly high (a
30\%--40\% enhancement) for their SMBH masses and luminosities (see also
Shemmer \& Netzer 2002; Shemmer et al.  2003), but still well below the
metallicities derived for most luminous quasars.

	It is difficult to compare our results directly to Shemmer \& Netzer
(2002) because they measure only \ion{N}{5}/\ion{C}{4} and
\ion{N}{5}/\ion{He}{2} ratios for a sample of individual objects.  However, it
does appear that the metallicity enhancement we derive as an average from
several line ratios is smaller than the Shemmer \& Netzer result based only on
\ion{N}{5}/\ion{C}{4} and \ion{N}{5}/\ion{He}{2}.

%	If SMBH mass is the fundamental parameter affecting BLR metallicity
%(Warner et al. 2003), it is reasonable to expect no trend between metallicity
%and $L/L_{edd}$ because there is no trend between $L/L_{edd}$ and SMBH mass.
%However, the NLS1 composite again exhibits unusual behavior.  The average
%SMBH mass and luminosity of the NLS1 composite are nearly two orders of
%magnitude lower than those of any of the $L/L_{edd}$ composites.  Therefore,
%if either of those quantities were the fundamental parameter affecting BLR
%metallicity, one would expect the NLS1 composite to have a significantly
%lower metallicity than the $L/L_{edd}$ composites.  On the contrary, Figure
%8 shows the NLS1 composite to have a metallicity comparable to that of the
%$L/L_{edd}$ composites:  much higher than would be expected for its SMBH mass
%and luminosity. 

\section{Discussion}

	There is evidence supporting super-Eddington accretion rates for
$\sim$27\% of the objects shown in Figures 1 and 4.  Previous studies have
also found quasars that appear to be accreting at super-Eddington rates
(e.g., Collin et al. 2002, Vestergaard 2004).  It is also worth noting that
changing the index on the $R_{BLR} - L$ relationship (see \S1) from 0.7 to
0.5 (e.g. Netzer \& Laor 1993; Shields et al. 2003) would actually increase
our estimates of $L/L_{edd}$, by as much as a factor of $\sim$ 4 in the
brightest quasars (see also Netzer 2003).
Super-Eddington accretion can be explained simply by non-spherically
symmetric accretion (Osterbrock 1989).  It has been suggested that
accretion disks with radiation-driven inhomogeneities could produce
luminosities up to 100 times Eddington (Begelman 2002; Wang 2003).

       Woo \& Urry (2002) find that only 9\%\ of their objects (21 out of
234, see their Fig. 7) have
$L/L_{edd} > 1$.  However, this may be due to the selection effect that their
sample contains only 9 objects with $L > 10^{47}$ ergs s$^{-1}$.  In our
sample, only 13\%\ of the objects with $L \leq 10^{47}$ ergs s$^{-1}$
have $L/L_{edd} > 1$ (compared to 8\%, 18 out of 225, in Woo \& Urry 2002),
but 41\%\ of the objects in our sample with $L > 10^{47}$ ergs s$^{-1}$
have super-Eddington ratios.

	The NLS1 composite spectrum exhibits many properties unlike
the $L/L_{edd}$ composite spectra.  The NLS1 composite has a much steeper
(``softer") UV continuum than the $L/L_{edd}$ composites,
consistent with studies finding that NLS1s generally have redder UV
continua than typical AGNs (e.g., Crenshaw et al. 2002; Constantin \& Shields
2003).  There is no
apparent trend between continuum shape and $L/L_{edd}$ (Table 1), but there
are trends between continuum shape and both SMBH mass (see Table 1; Warner et
al. 2003) and luminosity (Table 1; M. Dietrich, private comm.), such that
objects with lower SMBH masses and lower luminosities have steeper (``softer")
UV continua.  Therefore, the steeper UV continuum in the NLS1 composite may
be due to the NLS1 composite having a much lower average $M_{\rm SMBH}$ and
$L$ than the $L/L_{edd}$ composites.

	Interestingly, Figure 2 shows that the NLS1 composite
spectrum most strongly resembles the composite spectrum created from objects
with $L/L_{edd} > 2$.  It seems to fit at the top of Figure 2 and not where it
would be placed based on its average $L/L_{edd}$ of 0.67.  Based on this
similarity between NLS1s and quasars with high $L/L_{edd}$, it has been
suggested that high-redshift ($z \gtrsim 4$), high-luminosity, narrow-lined
quasars are analogs of NLS1s (Mathur 2000).  However, despite the similarity
in Figure 2 between the NLS1 composite and the $L/L_{edd} > 2$ composite,
the REWs of some emission lines in the two spectra can be quite different
(see Figure 7, Constantin \& Shields 2002).  Moreover, the NLS1s
do not fit the general trends in Figure 7 between emission line REW and
$L/L_{edd}$.

	If host galaxy mass, which correlates strongly with SMBH mass
(see \S1), is the fundamental parameter affecting BLR metallicity, it is
reasonable to expect no trend between metallicity and $L/L_{edd}$ because
there is no trend between $L/L_{edd}$ and SMBH mass in the $L/L_{edd}$
composites.  The results in Figures 10 and 11 confirm this expectation.  The
different $L/L_{edd}$ composites have similar average SMBH masses (Table 1)
and similar metallicities.  However, the NLS1 behavior is surprising.  They
have roughly the same metallicity as the $L/L_{edd}$ composites even though
their average SMBH mass and luminosity are almost two orders of magnitude
lower.  Figure 11 shows more directly that the NLS1s have slightly high
metallicities for their luminosities and SMBH masses (see also Shemmer \&
Netzer 2002; Shemmer et al. 2003).  If host galaxy mass, correlated with
$M_{\rm SMBH}$, is the main driver behind AGN metallicities (Figure 11 and
Warner et al. 2003), then clearly some other factor is enhancing the
metallicities in NLS1s.  The magnitude of the NLS1 enhancement is modest,
roughly 30\%--40\%.  Note, in particular, that the NLS1 metallicities are
still well below the values derived for most luminous (e.g., high-redshift)
quasars.

%The NLS1 composite, with an average SMBH mass and luminosity
%nearly two orders of magnitude lower than any of the $L/L_{edd}$ composites,
%may then be expected to have a lower metallicity than any of the $L/L_{edd}$
%composites.  However, Figure 10 shows that the NLS1 composite and the
%$L/L_{edd}$ composites have similar metallicities.  The NLS1s thus exhibit 
%exhibit high metallicities for their SMBH masses (see also Shemmer \& Netzer
%2002).  However, these metallicities are still much lower than the
%metallicities of high-redshift, high-luminosity quasars with broad
%emission-lines.  These quasars generally fall at the high end of a trend
%between metallicity and SMBH mass (see Warner et al. 2003), but the NLS1s lie
%above this trend.  Therefore, if SMBH mass is related to the driving parameter
%affecting BLR metallicity, there must be some additional effect that is
%enhancing the metallicity in NLS1s.

	Shemmer \& Netzer (2002) suggest that high $L/L_{edd}$ is driving the
high metallicities in NLS1s.  However, we have shown that there
is no correlation between metallicity and $L/L_{edd}$ (Figures 10 and 11).
In addition, luminous narrow-lined quasars with the highest values of
$L/L_{edd}$ (such as BR2248-1242, Warner et al. 2002) do {\it not} have
high metallicities for their SMBH masses.  We conclude that the additional
parameter affecting NLS1 metallicities is not related to $L/L_{edd}$.

%the luminous quasars with the highest $L/L_{edd}$ (e.g., those with narrow
%\ion{C}{4} emission-lines like BR2248-1242) do not have high metallicities
%for their SMBH masses and luminosities.  We therefore
%conclude that the additional effect enhancing the metallicity in NLS1s is
%not related to $L/L_{edd}$.

	Constantin \& Shields (2003) suggest that the result for NLS1s having
lower metallicities than high-redshift quasars runs counter to the hypothesis
that these two groups of objects are in a similar early evolutionary phase.
However, because the high-redshift quasars have higher SMBH masses and
luminosities than NLS1s, they would naturally be expected to have higher
metallicities.  If both types of objects are in similar early evolutionary
phases, and/or reside in young or rejuvenated host galaxies and the
metallicities of both types of objects are thus enhanced by a similar amount,
then the high-redshift quasars would still exhibit a higher metallicity than
the NLS1s due to their higher SMBH masses.  Thus, we conclude that more
information is necessary to determine if either NLS1s or the high-redshift
quasars (or both) are preferentially young objects.

\section{Summary \& Conclusions}
	We have examined a large sample of 578 AGNs that spans five orders of
magnitude in SMBH mass, seven orders of magnitude in luminosity, and a redshift
range from $0 \leq z \leq 5$.  We estimate SMBH masses using the virial
theorem and formulae given in Kaspi et al. (2000), and then derive Eddington
ratios.  To improve the signal-to-noise ratio and average over object-to-object
variations, we calculate composite spectra for different ranges in $L/L_{edd}$.
We include a composite spectrum of a sample of 26 NLS1s for comparative
analysis.  Our main results are as follows.

	1) We find that a large fraction (27\%) of the objects in our sample
have $L/L_{edd} > 1$.  These super-Eddington ratios may be explained by
non-spherically symmetric accretion.  While NLS1s generally show high
Eddington ratios for their luminosities, the objects with the highest
$L/L_{edd}$ are high-luminosity, narrow-lined quasars.

	2) There is no trend between $L/L_{edd}$ and either redshift or
SMBH mass.  $L/L_{edd}$ does correlate positively with luminosity and
negatively with FWHM(\ion{C}{4}), but these trends may be attributed largely
to our derivation of $L/L_{edd}$ from these quantities (see Equation 4).

	3) There is no trend between the shape of the UV continuum and
$L/L_{edd}$.  The NLS1 composite has a much steeper (softer) continuum
than the $L/L_{edd}$ composites.  This is consistent with a trend between
continuum shape and SMBH mass.	

	4) The composite spectra sorted by $L/L_{edd}$ exhibit an unusual
emission-line behavior: nearly constant peak heights and decreasing FWHMs
with increasing $L/L_{edd}$ (Figure 2).  The origins of this behavior are not
understood, but it is in marked contrast to the emission-line behaviors in
composite spectra sorted by luminosity, SMBH mass, and FWHM(\ion{C}{4})
(Figure 8), all of which clearly show trends in the line REWs analagous to
the Baldwin Effect.

	5) The composite spectra show no trend between $L/L_{edd}$ and
metallicity (Figure 10).  This is consistent with SMBH mass being related to
the fundamental parameter affecting BLR metallicity (Warner et al. 2003).

	6) The NLS1 composite exhibits several unusual behaviors.  It generally
does not fit the trends between emission line REWs and $L/L_{edd}$ as
defined by the $L/L_{edd}$ composites.  It also has a metallicity that is 
slightly high for its average SMBH mass and luminosity, although still well
below the high metallicities exhibited by the most luminous quasars with
the most massive central SMBHs.  The quasars with the highest $L/L_{edd}$,
high-luminosity quasars with narrow \ion{C}{4} emission, do not have high
metallicities for their SMBH masses and luminosities.  Our earlier work
(Warner et al. 2003) is consistent with the theory that host galaxy mass,
correlated with SMBH mass (and AGN luminosity), is the fundamental parameter
affecting BLR metallicity.  We conclude that i) there must be some secondary 
effect enhancing the metallicity in NLS1s, and ii) this secondary effect
is not related to $L/L_{edd}$.

\noindent {\it Acknowledgements:} We are very grateful to Marianne Vestergaard
for providing the UV Fe emission template for this study, and to Fred Chaffee,
Anca Constantin, Craig Foltz, Vesa Junkkarinen, and Joe Shields for their
direct participation in reducing or acquiring some of the ground-based
spectra.  We acknowledge financial support from the NSF
via grant AST99-84040 and NASA via grant NAG5-3234. 

%\appendix

\newpage
\begin{figure}
\vbox{
\centerline{
\psfig{figure=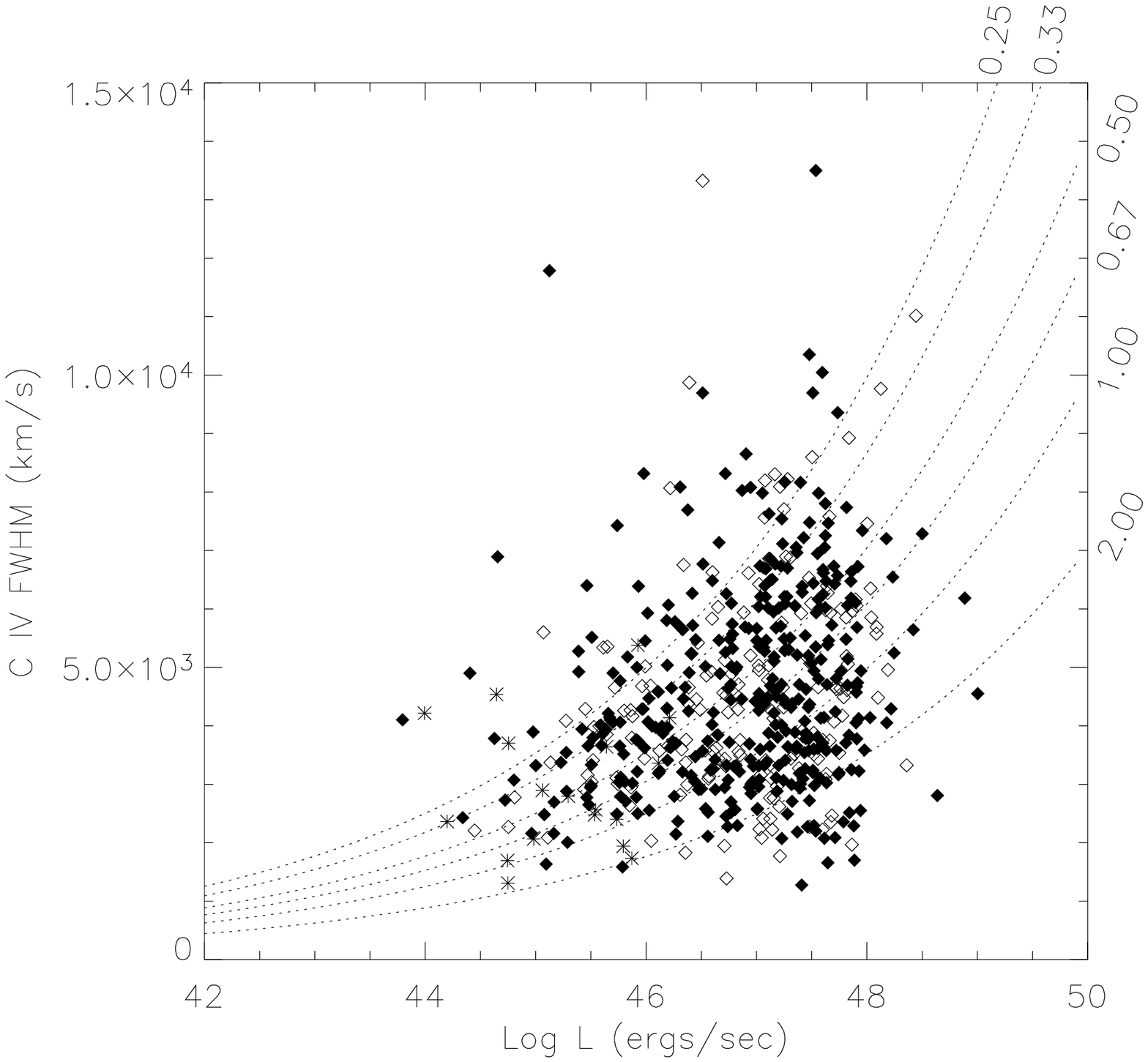}
}
\vskip 14pt
\caption
{ Distribution of the two measured quantities.  The dotted lines represent
constant $L/L_{edd}$ ratios, with values indicated at the right.  We created
composite spectra for each of the seven ranges in Eddington ratio shown above.
The filled diamonds represent radio-quiet quasars, the open diamonds represent
radio-loud quasars, and the asterisks represent NLS1s.
}
}
\end{figure}

\begin{figure}
\vbox{
\centerline{
\psfig{figure=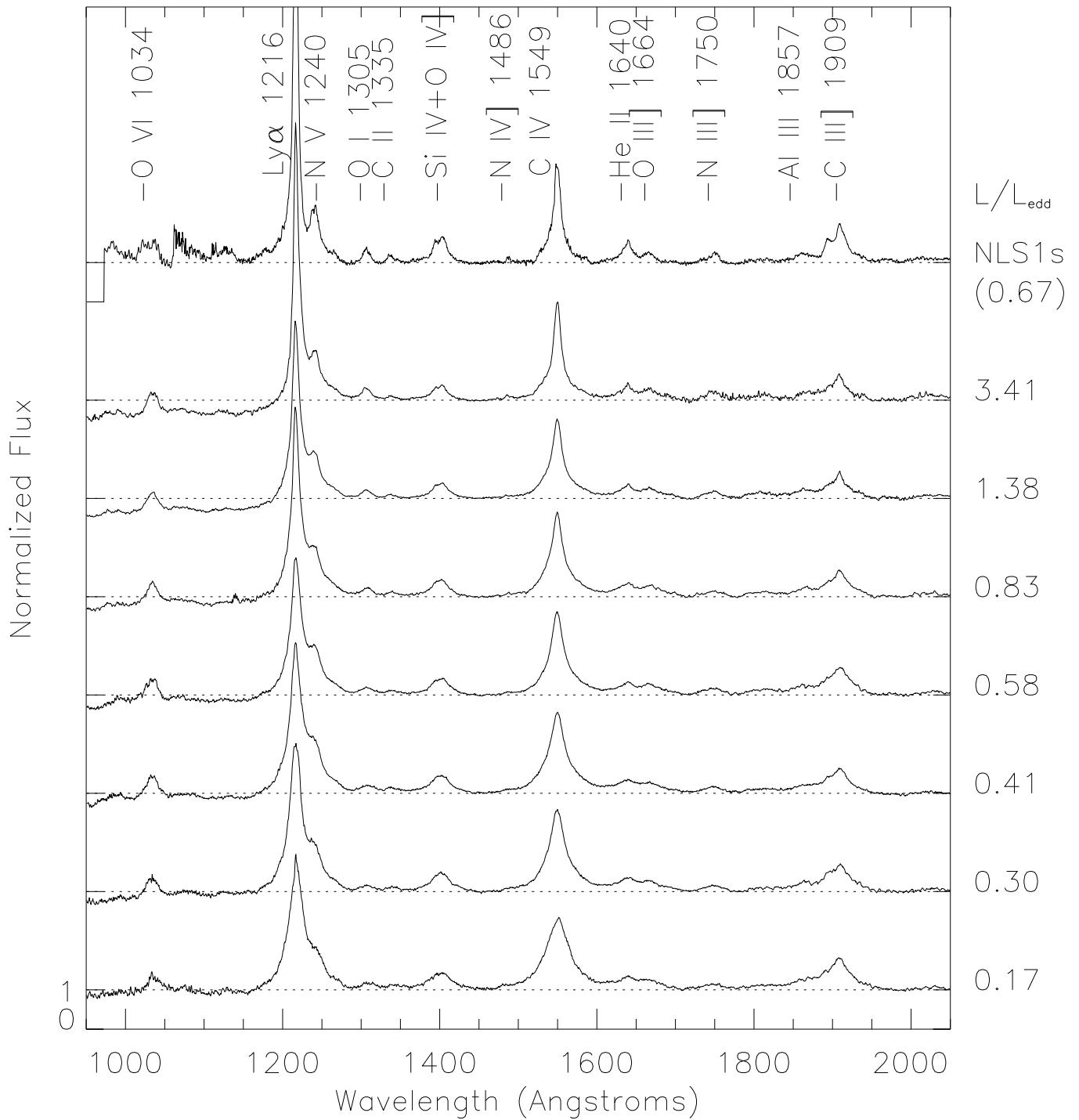}
}
\vskip 14pt
\caption
{
Composite spectra, sorted by $L/L_{edd}$, with the mean $L/L_{edd}$ values
indicated at the right.  The top spectrum is a composite spectrum created
only from objects identified in the literature as NLS1s.
The horizontal dashed lines and tick marks indicate the normalized continuum
levels for each composite spectrum.  The scaled height of the normalized
continua above zero flux, as indicated for the bottom spectrum, is the same
for all spectra plotted.
}
}
\end{figure}

\begin{figure}
\vbox{
\centerline{
\psfig{figure=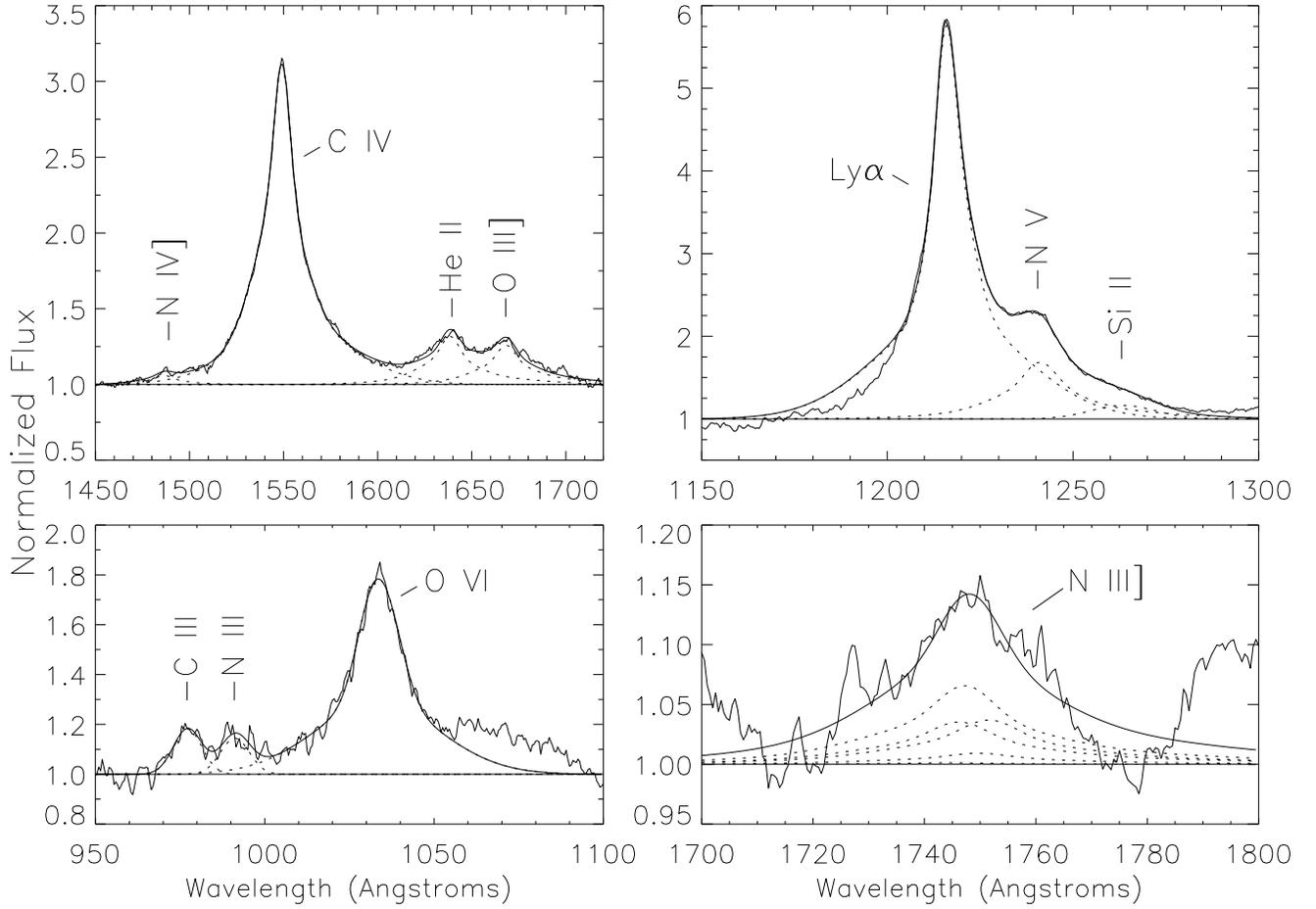}
}
\vskip 14pt
\caption
{
Multi-component Gaussian fits (smooth solid curves) to observed emission
lines (jagged solid curves) for a composite spectrum of $<L/L_{edd}> \sim
0.58$.  The continuum is normalized by a power law continuum fit.  The
continuum (at unity) and composite line fits are plotted.  The individual
components of the fits are shown for the fit to \ion{N}{3}] as dotted lines.
In the other panels, the dotted lines represent the sum contribution of
each labelled emission line.  There is an unidentified bump at $\sim$1070
\AA\ that is not attributed to \ion{O}{6} (see Hamann et al. 1998).
}
}
\end{figure}

\begin{figure}
\vbox{
\centerline{
\psfig{figure=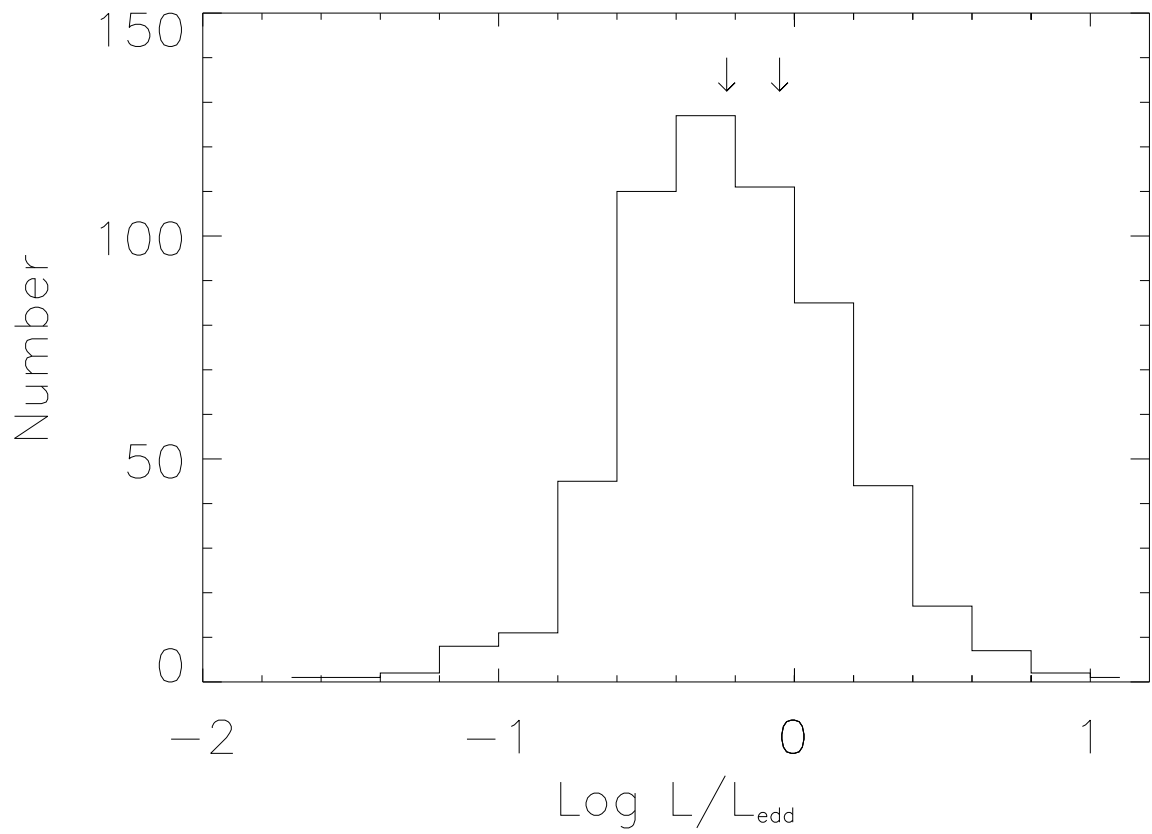}
}
\vskip 14pt
\caption
{
Distribution of estimated Eddington ratio, $L/L_{edd}$.  The two arrows are
the median (0.59) and mean (0.89) $L/L_{edd}$ of the entire sample.  About
27\% of the objects have $L/L_{edd} > 1$.
}
}
\end{figure}

\begin{figure}
\vbox{
\centerline{
\psfig{figure=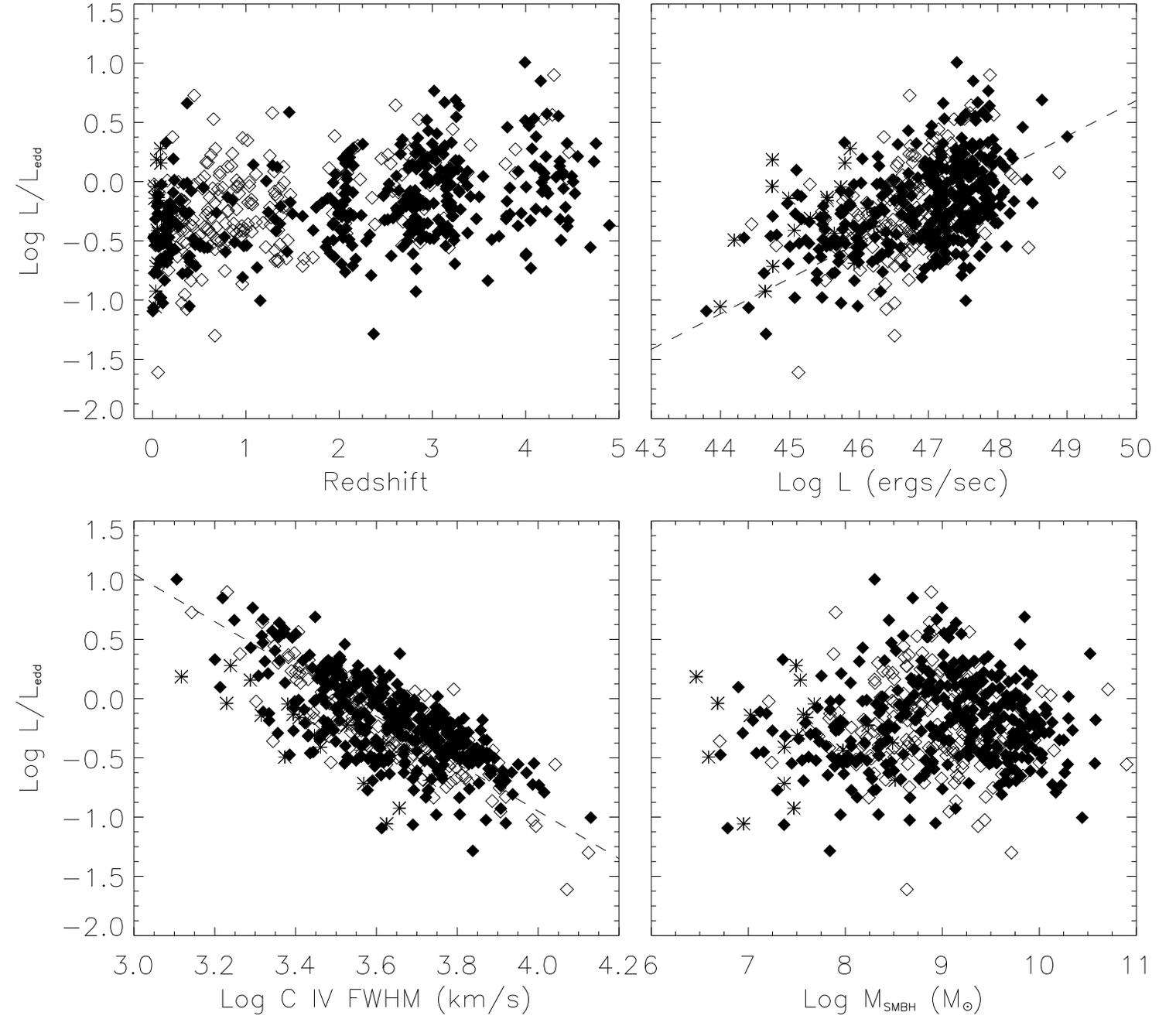}
}
\vskip 14pt
\caption
{Symbols as in Fig. 1.  The dashed lines represent slopes from the
$L/L_{edd}$ equation (not fits).  The positive correlation with $L$
and negative correlation with FWHM(\ion{C}{4}) may be attributed largely to
our derivation of $L/L_{edd}$ from these quantities (see Equation 4).
}
}
\end{figure}

\begin{figure}
\vbox{
\centerline{
\psfig{figure=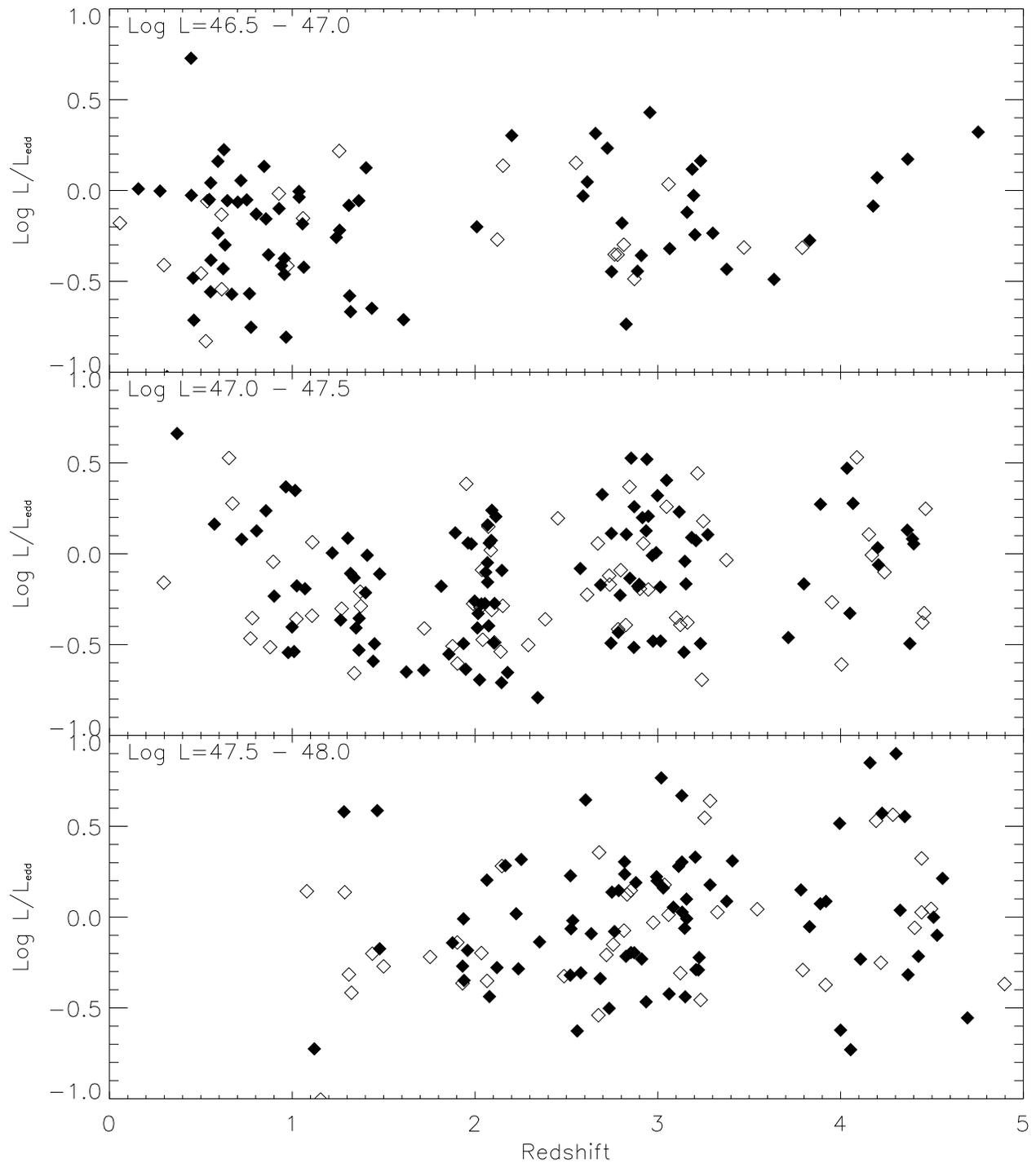}
}
\vskip 14pt
\vskip 18pt
\caption
{
Symbols as in Fig. 1.
In bins of narrow luminosity ranges, there is no apparent trend between
$L/L_{edd}$ and redshift.  Our sample, though, does seem to have a selection
effect for more higher luminosity objects at higher redshifts.  This, combined
with the trend for higher luminosity objects to have higher $L/L_{edd}$,
explains the slight trend that seems to exist in Fig. 5 between $L/L_{edd}$
and redshift.
}
}
\end{figure}

\begin{figure}
\vbox{
\centerline{
\psfig{figure=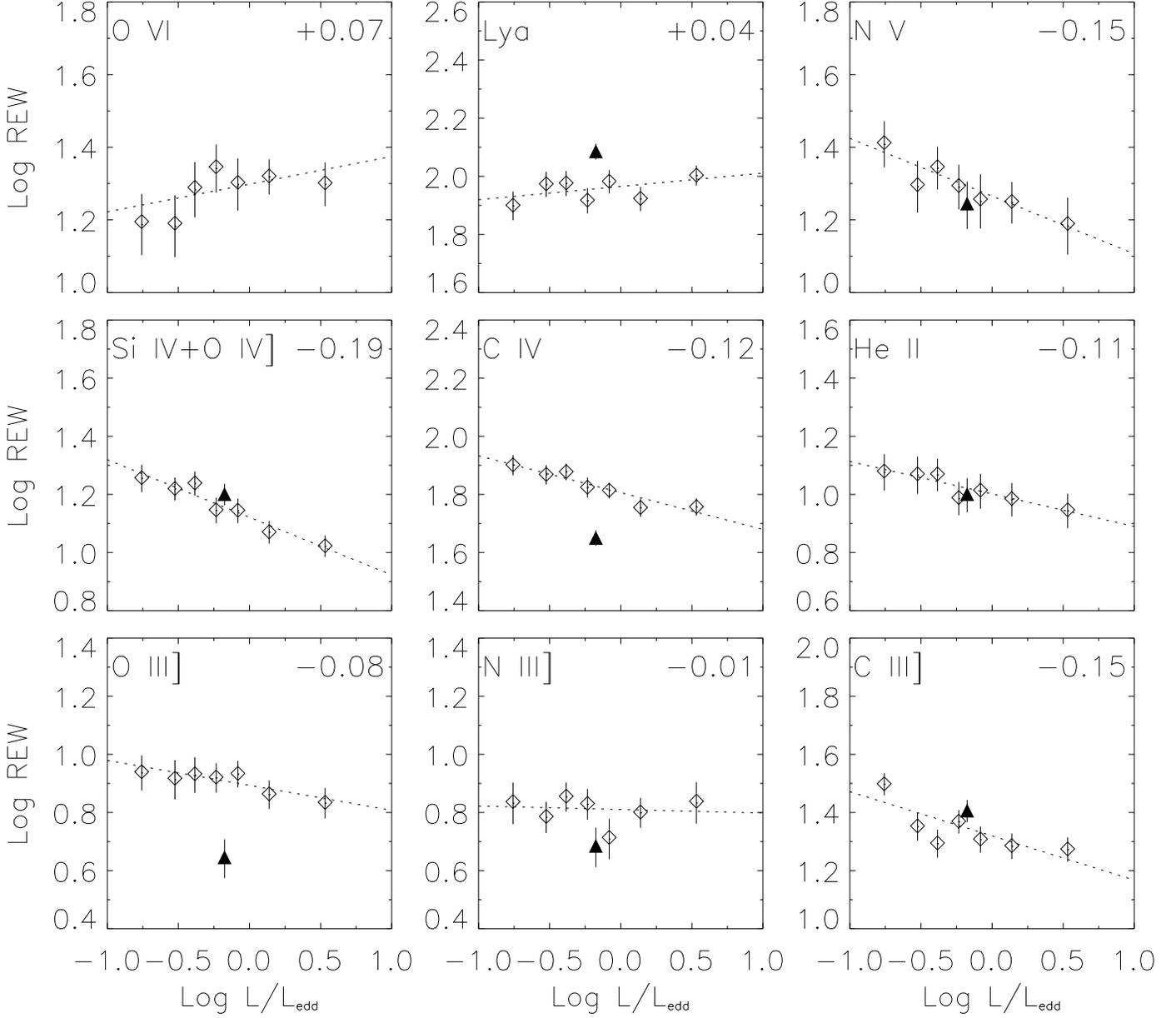}
}
\vskip 14pt
\caption
{
The diamonds represent the $L/L_{edd}$ composites and the triangle represents
the NLS1 composite.  REWs are in angstroms.  The dotted lines are linear fits
to the $L/L_{edd}$ composites, with slopes given in the upper right.  The
uncertainties shown are the $1\sigma$ standard deviations in the REWs based on
repeated estimates with the continuum drawn at different levels (see \S3).
}
}
\end{figure}

\begin{figure}
\vbox{
\centerline{
\psfig{figure=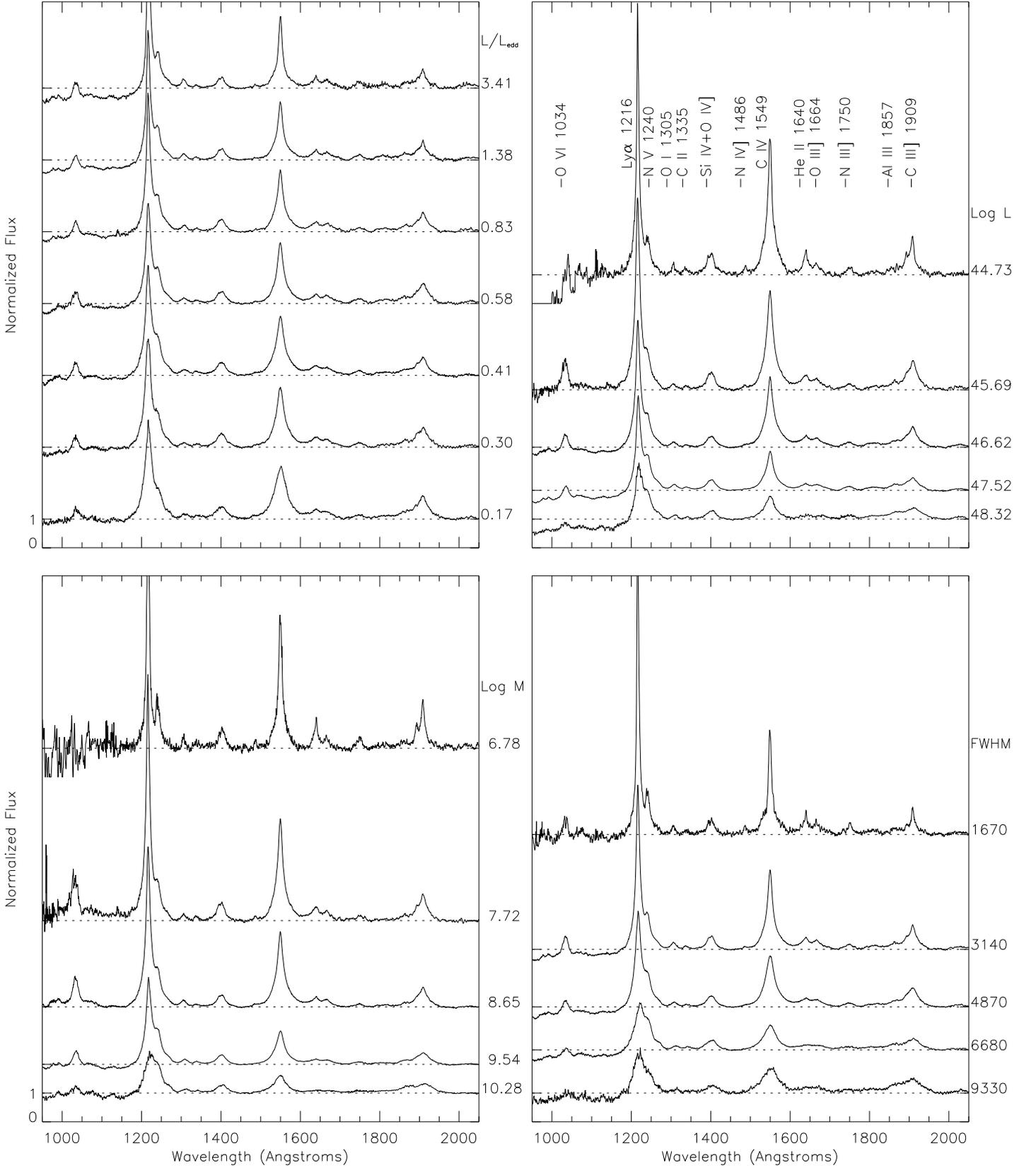}
}
\vskip 14pt
\vskip 18pt
\caption
{
Normalized composite spectra sorted by $L/L_{edd}$ (top left, identical to
Fig. 2 without the NLS1 composite) are compared
to normalized composite spectra computed from different ranges in bolometric
luminosity (top right, in units of ergs/sec), SMBH mass (bottom left, in
\Msun), and FWHM(\ion{C}{4}) (bottom right, in km/sec).
Dashed lines and tick marks as in Fig. 2.  The $L$, SMBH
mass, and FWHM composite spectra all clearly show trends for decreasing
line equivalent widths and peak heights as luminosity, SMBH mass, and
FWHM(\ion{C}{4}) increase.  The composite spectra sorted by FWHM(\ion{C}{4})
even exhibit this trend despite spanning less than half an order of magnitude
in average luminosity, a much narrower range than in the $L/L_{edd}$
composites (see Table 1).  However, the $L/L_{edd}$ composites show
nearly constant peak heights and decreasing FWHMs with increasing $L/L_{edd}$.
}
}
\end{figure}

\begin{figure}
\vbox{
\centerline{
\psfig{figure=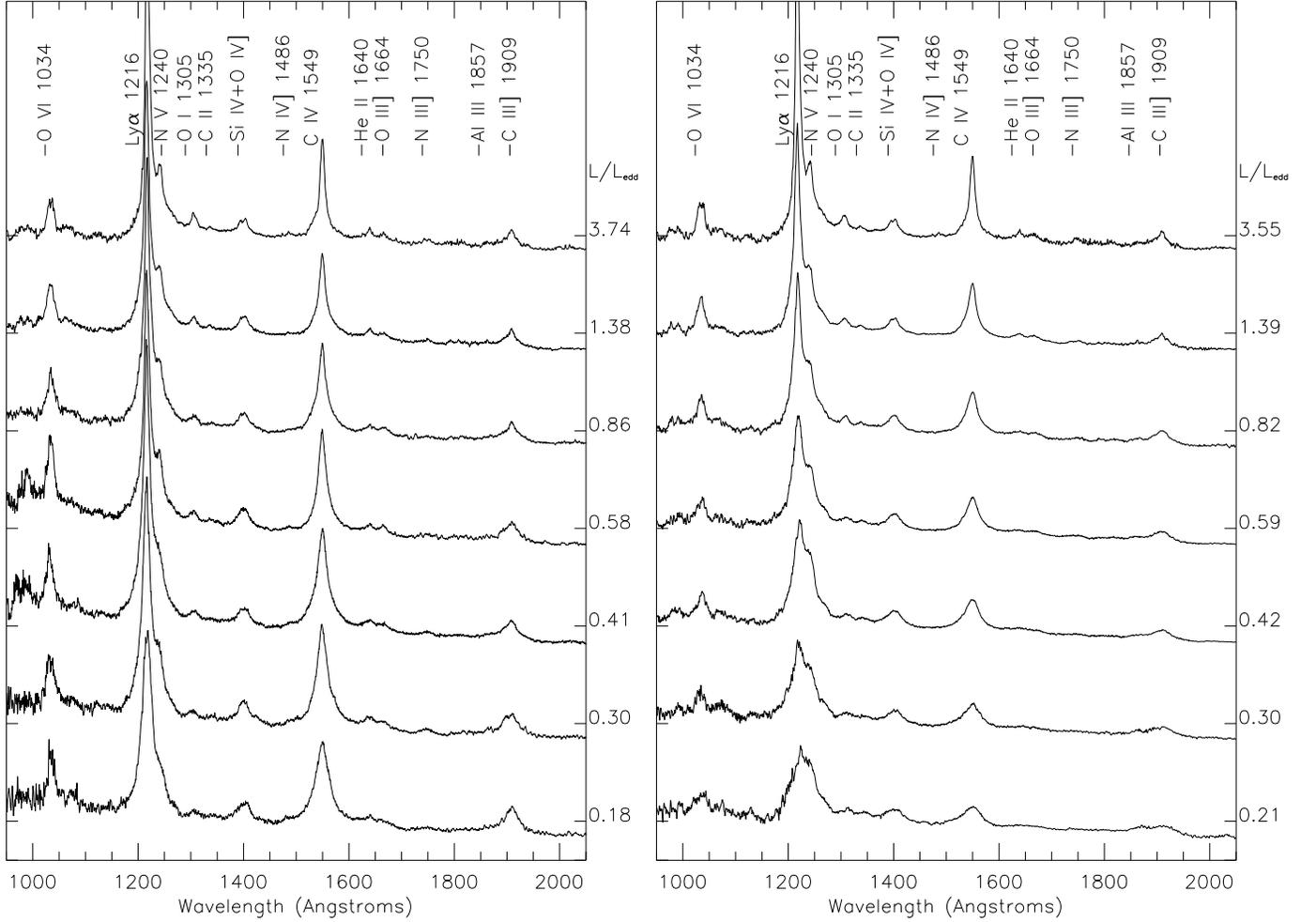}
}
\vskip 14pt
\vskip 18pt
\caption
{
Composite spectra sorted by $L/L_{edd}$ for a narrow range in SMBH mass of
$10^8 \Msun < M_{\rm SMBH} < 10^9 \Msun$ (left) are compared to ones sorted
by $L/L_{edd}$ for narrow range in luminosity of  $10^{47}$ ergs/s $< L <
10^{48}$ ergs/s (right).  The spectra are shown prior to 
normalization to the continuum.  The composite spectra at nearly constant
$M_{\rm SMBH}$ show the same emission-line behavior as the $L/L_{edd}$
composites created from our entire sample: nearly constant peak heights and
decreasing FWHMs with increasing $L/L_{edd}$.  In contrast, the composite
spectra at nearly constant $L$ do not share this behavior and show a trend
similar to the Baldwin Effect: decreasing peak heights and equivalent
widths with increasing $M_{\rm SMBH}$.
}
}
\end{figure}

\begin{figure}
\vbox{
\centerline{
\psfig{figure=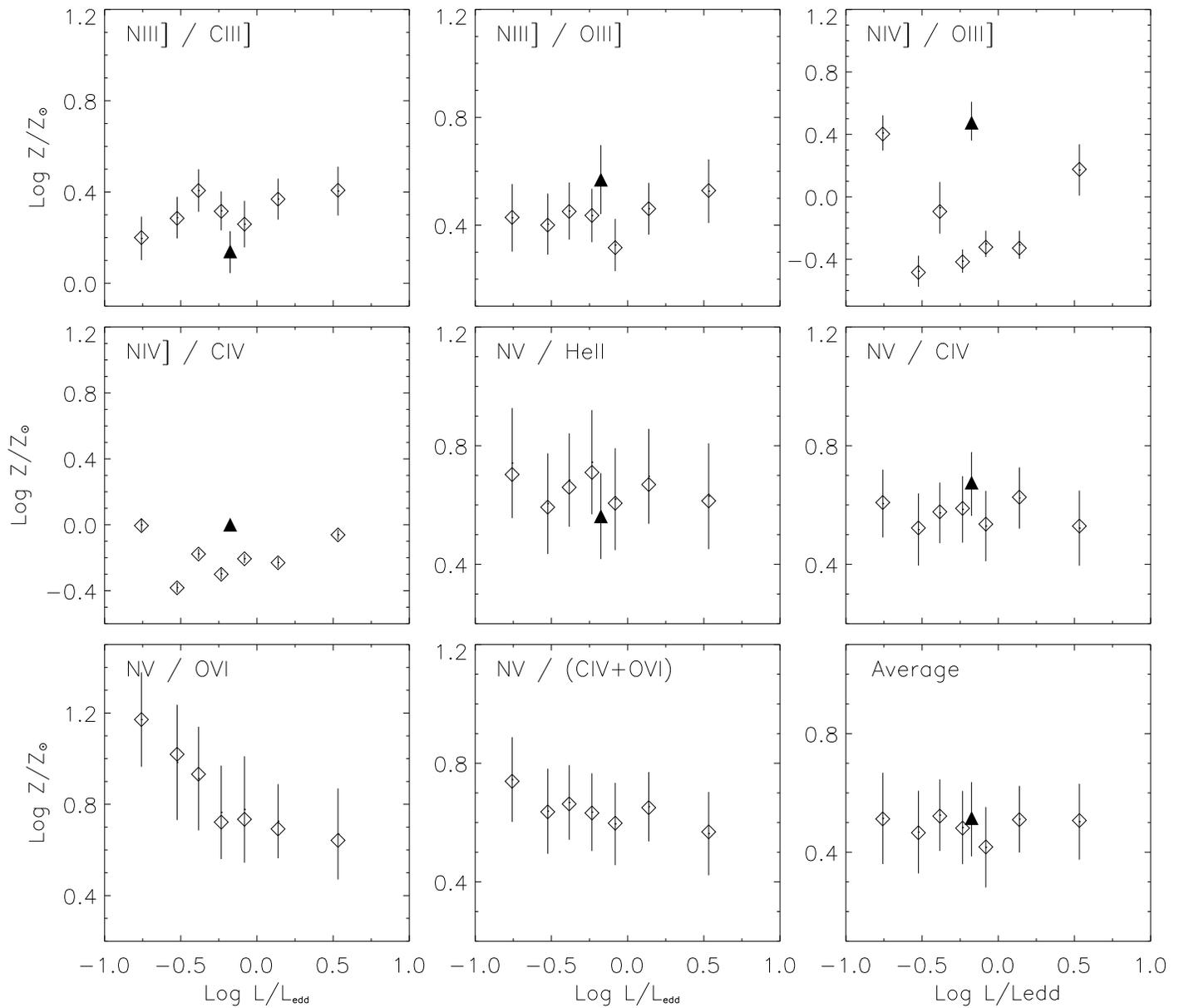}
}
\vskip 18pt
\caption
{
Metallicities derived from comparisons of different line ratio diagnostics
to theoretical results from Figure 5 in Hamann et al. (2002) are shown as
a function of $L/L_{edd}$.  The average includes \ion{N}{3}]/\ion{C}{3}],
\ion{N}{3}]/\ion{O}{3}], and \ion{N}{5}/(\ion{C}{4}+\ion{O}{6}).  Symbols
as in Fig. 7. 
}
}
\end{figure}

\begin{figure}
\vbox{
\centerline{
\psfig{figure=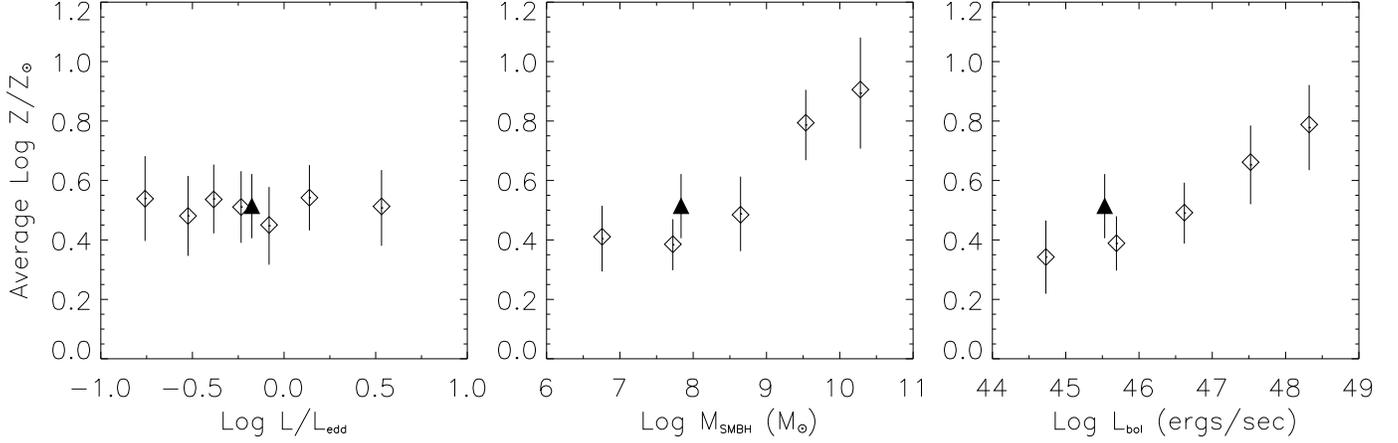}
}
\vskip 18pt
\caption
{
Symbols as in Fig. 7.  Our best estimates of the overall metallicity of each
spectrum are shown for the $L/L_{edd}$, SMBH mass, and luminosity composite
spectra.  The NLS1 composite, plotted in all three panels as a filled triangle,
has a metallicity that is slightly high for its SMBH mass and luminosity, but
still well below the metallicities of most luminous quasars.
}
}
\end{figure}

%\begin{table}
%\tablewidth{485.22502pt}
%\begin{center}
%\caption{Composite Parameters}
%\vspace*{0.1in}
%\begin{tabular}{cccccccc}
%\hline
%\hline
%\noalign{\vskip 4pt}
%$L/L_{edd}$ & $\alpha$ & \# Objects & $z$ & Log $L$ & FWHM(\ion{C}{4}) & Log $M_{\rm BH}$\\
%\noalign{\vskip 2pt}
% & & at \ion{C}{4} & & [ergs s$^{-1}$] & [km s$^{-1}$] & [\Msun] \\
%\noalign{\vskip 2pt}
%\hline
%\noalign{\vskip 4pt}
%0.17 & -0.65 & 64 & 1.06 & 46.84 & 6900 & 9.48\\
%\noalign{\vskip 4pt}
%0.30 & -0.56 & 62 & 1.17 & 47.17 & 5600 & 9.60\\
%\noalign{\vskip 4pt}
%0.41 & -0.49 & 106 & 1.62 & 47.16 & 5100 & 9.44\\
%\noalign{\vskip 4pt}
%0.58 & -0.40 & 85 & 1.83 & 47.43 & 4600 & 9.56\\
%\noalign{\vskip 4pt}
%0.83 & -0.60 & 90 & 1.93 & 47.37 & 3900 & 9.36\\
%\noalign{\vskip 4pt}
%1.38 & -0.33 & 110 & 2.55 & 47.63 & 3400 & 9.42\\
%\noalign{\vskip 4pt}
%3.41 & -0.54 & 45 & 2.86 & 47.84 & 2400 & 9.26\\
%\noalign{\vskip 4pt}
%0.67 & -0.93 & 18 & 0.06 & 45.53 & 2900 & 7.83\\
%\noalign{\vskip 2pt}
%\hline
%\end{tabular}
%\end{center}
%\end{table}

\begin{deluxetable}{ccccccc}
\tabletypesize{\normalsize}
\tablewidth{340.22502pt}
\tablecaption{Composite Parameters}
\tablehead{
\colhead{$L/L_{edd}$} & 
\colhead{$\alpha$} & 
\colhead{\# Objects} &
\colhead{$z$} & 
\colhead{Log $L$} & 
\colhead{FWHM(\ion{C}{4})} & 
\colhead{Log $M_{\rm SMBH}$} \\
\colhead{} & 
\colhead{} & 
\colhead{at \ion{C}{4}} & 
\colhead{} & 
\colhead{[ergs s$^{-1}$]} & 
\colhead{[km s$^{-1}$]} & 
\colhead{[\Msun]}
}
\tablecolumns{7}
\startdata
\multicolumn{7}{c}{-----------------------------------  $L/L_{edd}$ Composites  -----------------------------------}\\
\noalign{\vskip 2 pt}
0.17 & -0.65 & 64 & 1.06 & 46.84 & 6900 & 9.48\\
\noalign{\vskip 2pt}
0.30 & -0.56 & 62 & 1.17 & 47.17 & 5600 & 9.60\\
\noalign{\vskip 2pt}
0.41 & -0.49 & 106 & 1.62 & 47.16 & 5100 & 9.44\\
\noalign{\vskip 2pt}
0.58 & -0.40 & 85 & 1.83 & 47.43 & 4600 & 9.56\\
\noalign{\vskip 2pt}
0.83 & -0.60 & 90 & 1.93 & 47.37 & 3900 & 9.36\\
\noalign{\vskip 2pt}
1.38 & -0.33 & 110 & 2.55 & 47.63 & 3400 & 9.42\\
\noalign{\vskip 2pt}
3.41 & -0.54 & 45 & 2.86 & 47.84 & 2400 & 9.26\\
\noalign{\vskip 2pt}
\multicolumn{7}{c}{-----------------------------------------  NLS1s  -----------------------------------------}\\
\noalign{\vskip 2pt}
0.67 & -0.93 & 18 & 0.06 & 45.53 & 2900 & 7.83\\
%\enddata
%\end{deluxetable}

%\begin{deluxetable}{ccccccc}
%\tabletypesize{\normalsize}
%\tablewidth{340.22502pt}
%\tablecaption{$L$, $M_{\rm SMBH}$, and FWHM(\ion{C}{4}) Composite Parameters}
%\tablehead{
%\colhead{$L/L_{edd}$} &
%\colhead{$\alpha$} &
%\colhead{\# Objects} &
%\colhead{$z$} &
%\colhead{Log $L$} &
%\colhead{FWHM(\ion{C}{4})} &
%\colhead{Log $M_{\rm SMBH}$} \\
%\colhead{} &
%\colhead{} &
%\colhead{at \ion{C}{4}} &
%\colhead{} &
%\colhead{[ergs s$^{-1}$]} &
%\colhead{[km s$^{-1}$]} &
%\colhead{[\Msun]}
%}
%\tablecolumns{7}
%\startdata
\noalign{\vskip 2pt}
\multicolumn{7}{c}{------------------------------------  $L$ Composites  ------------------------------------}\\
\noalign{\vskip 2pt}
0.43 & -0.73 & 15 & 0.57 & 44.73 & 3100 & 7.25\\
0.46 & -0.66 & 90 & 0.37 & 45.69 & 3800 & 8.10\\
0.69 & -0.36 & 155 & 1.39 & 46.62 & 4400 & 8.87\\
1.15 & -0.47 & 278 & 2.63 & 47.52 & 4900 & 9.59\\
1.21 & -0.24 & 20 & 3.01 & 48.32 & 5900 & 10.30\\
\noalign{\vskip 2pt}
\multicolumn{7}{c}{-----------------------------------  $M_{\rm SMBH}$ Composites  -----------------------------------}\\
\noalign{\vskip 2pt}
0.61 & -0.98 & 7 & 0.56 & 44.62 & 2500 & 6.78\\
0.69 & -0.90 & 61 & 0.46 & 45.68 & 3000 & 7.27\\
1.08 & -0.71 & 198 & 1.43 & 46.84 & 3800 & 8.65\\
0.84 & -0.59 & 261 & 2.46 & 47.52 & 5200 & 9.54\\
0.58 & -0.57 & 34 & 2.84 & 48.17 & 7500 & 10.28\\
\noalign{\vskip 2pt}
\multicolumn{7}{c}{--------------------------------  FWHM(\ion{C}{4}) Composites  --------------------------------}\\
\noalign{\vskip 2pt}
3.94 & -0.91 & 14 & 1.64 & 47.25 & 1700 & 8.35\\
1.24 & -0.48 & 237 & 1.71 & 47.22 & 3100 & 8.87\\
0.58 & -0.38 & 195 & 1.86 & 47.48 & 4900 & 9.43\\
0.38 & -0.42 & 91 & 2.29 & 47.65 & 6700 & 9.84\\
0.17 & -0.47 & 25 & 1.89 & 47.52 & 9300 & 10.07\\
\enddata
\end{deluxetable}

\begin{deluxetable}{lccccccccccccccc}
\rotate
\tabletypesize{\small}
%\tabletypesize{\tiny}
\tablecaption{Emission Line Data}
\tablewidth{565.22502pt}
%\tablewidth{645.22502pt}
\tablehead{
\colhead{$L/L_{edd}$} &
\colhead{Property} &
\colhead{\ion{O}{6} $\lambda 1034$} &
\colhead{\Lya\ $\lambda 1216$} &
\colhead{\ion{N}{5} $\lambda 1240$} &
\colhead{\ion{C}{4} $\lambda 1549$} &
\colhead{\ion{He}{2} $\lambda 1640$} &
\colhead{\ion{O}{3}] $\lambda 1665$} &
\colhead{\ion{N}{3}] $\lambda 1750$} &
\colhead{\ion{C}{3}] $\lambda 1909$} &
}
\tablecolumns{16}
\startdata
%\ion{C}{4}  $\lambda 1549$ & 0.556 & 87 & 2100 & 0.547 & 94 & 3100
%& 0.597 & 80 & 3800 & 0.462 & 41 & 4700 & 0.500 & 28 & 6500\\
0.17 & Flux/\Lya\ & 0.21 & 1.00 & 0.32 & 0.75 & 0.10 & 0.07 & 0.06 &0.22\\
     & REW$^{a}$ & 16 & 80 & 26 & 80 & 12 & 9 & 7 & 32\\
    & FWHM$^{b}$ & 5500 & 4700 & 6900 & 6800 & 6800 & 6800 & 6900 & 4900\\[3pt]
0.30 & Flux/\Lya\ & 0.17 & 1.00 & 0.21 & 0.57 & 0.08 & 0.06&0.04&0.13\\
     & REW$^{a}$ & 16 & 94 & 20 & 74 & 12 & 8 & 6 & 23\\
    & FWHM$^{b}$ & 5200 & 4300 & 5100 & 4800 & 4800 & 4800 & 5000 & 5000\\[3pt]
0.41 & Flux/\Lya\ & 0.21 & 1.00 & 0.23 & 0.57 & 0.08 & 0.06&0.05&0.11\\
     & REW$^{a}$ & 19 & 95 & 22 & 76 & 12 & 9 & 7 & 20\\
    & FWHM$^{b}$ & 6200 & 4600 & 5300 & 4900 & 4900 & 4900 & 5100 & 4300\\[3pt]
0.58 & Flux/\Lya\ & 0.28 & 1.00 & 0.24 & 0.57 & 0.08 & 0.06&0.05&0.14\\
     & REW$^{a}$ & 22 & 83 & 20 & 67 & 10 & 8 & 7 & 23\\
    & FWHM$^{b}$ & 6500 & 3800 & 4300 & 4000 & 4000 & 4000 & 4200 & 4900\\[3pt]
0.83 & Flux/\Lya\ & 0.22 & 1.000 & 0.19 & 0.50 & 0.07 & 0.06&0.03&0.12\\
     & REW$^{a}$ & 20 & 96 & 18 & 65 & 10 & 9 & 5 & 20\\
    & FWHM$^{b}$ & 5200 & 3000 & 4300 & 3700 & 3700 & 3700 & 4000 & 4000\\[3pt]
1.38 & Flux/\Lya\ & 0.26 & 1.00 & 0.21 & 0.47 & 0.07 & 0.05&0.04&0.11\\
     & REW$^{a}$ & 21 & 84 & 18 & 57 & 10 & 7 & 6 & 19\\
    & FWHM$^{b}$ & 5500 & 2800 & 3900 & 3300 & 3300 & 3300 & 3600 & 3500\\[3pt]
3.41 & Flux/\Lya\ & 0.21 & 1.00 & 0.15 & 0.41 & 0.06 & 0.04&0.04&0.10\\
     & REW$^{a}$ & 20 & 100 & 15 & 57 & 9 & 7 & 7 & 19\\
    & FWHM$^{b}$ & 5100 & 2100 & 3000 & 2500 & 2500 & 2500 & 2800 & 3100\\[3pt]
0.67 & Flux/\Lya\ & -- & 1.00 & 0.14 & 0.29 & 0.06 & 0.03 & 0.03&0.13\\
(NLS1s) & REW$^{a}$ & -- & 122 & 18 & 45 & 10 & 4 & 5 & 25\\
    & FWHM$^{b}$ & -- & 2100 & 2800 & 2200 & 2200 & 2200 & 2700 & 3100\\
\enddata
\tablenotetext{a}{In units of \AA}
\tablenotetext{b}{In units of km~s$^{-1}$}
\end{deluxetable}


\begin{references}
%\reference{} Baldwin, J. A. 1977, \apj, 214, 679
%\reference{} Baldwin, J. A., Hamann, F., Korista, K. T., Ferland, G. J., Dietrich, M., \& Warner, C. 2003, \apj, 583, 649 
\reference{} Begelman, M. C. 2002, \apjl, 568, L97
%\reference{} Bender, R., Burnstein, D., \& Faber, S. M. 1993, \apj, 411, 153
\reference{} Bettoni, D., Falomo, R., Fasano, G., \& Govoni, F. 2003, \aap, 399, 869
\reference{} Bischof, O. B. \& Becker, R. H. 1997, \aj, 113, 2000
\reference{} Boroson, T. A. 2002, \apj, 565, 78
\reference{} Boroson, T. A. 2003, \apj, 585, 647
\reference{} Boroson, T. A. \& Green, R. F. 1992, \apjs, 80, 109
\reference{} Brotherton, M. S. et al. 2001, \apj, 546, 775
\reference{} Carroll, S. M., Press, W. H., \& Turner, E. L. 1992, \araa, 30, 499
\reference{} Collin, S. et al. 2002, \aap, 388, 771
%\reference{} Constantin, A., Shields, J. C., Hamann, F., Foltz, C., \& Chaffee, F. 2001, \apj, 565, 50 
\reference{} Constantin, A. \& Shields, J. C. 2003, \pasp, 115, 592
\reference{} Corbett, E. A. et al. 2003, \mnras, 343, 705 
\reference{} Crenshaw, D. M. et al. 2002, \apj, 566, 187
\reference{} Croom, S. M. et al. 2002, \mnras, 337, 275 
\reference{} Dietrich, M. et al. 1999, \aap, 352, L1
\reference{} Dietrich, M., Hamann, F. et al. 2002, \apj, 581, 912 
%\reference{} Dietrich, M. et al. 2003a, \aap, 398, 891
%\reference{} Dietrich, M., Hamann, F. et al. 2003b, \apj, 589, in press
\reference{} Elvis, M. et al. 1994, \apjs, 95, 1
\reference{} Erwin, P., Graham, A. W., \& Caon, N. 2002, astro-ph/0212335
%\reference{} Espey, B. R., Lanzetta, K. M., \& Turnshek, D. A. 1993, \baas, 25, 1448
%\reference{} Faber, S. M. 1973, \apj, 179, 731 
\reference{} Ferland, G. J. et al. 1996, \apj, 461, 683
%\reference{} Ferland, G. J. et al. 1998, \pasp, 110, 761
\reference{} Ferrarese, L. \& Merritt, D. 2000, \apjl, 539, L9
%\reference{} Ferrarese, L. et al. 2001, \apjl, 555, L79
%\reference{} Ferrarese, L. \& Merritt, D. 2002, astro-ph/0206222
\reference{} Gebhardt, K. et al. 2000, \apjl, 539, L13
%\reference{} Gebhardt, K. et al. 2000b, \apjl, 543, L5 
%\reference{} Gnedin, N. Y. \& Ostriker, J. P. 1997, \apj, 486, 581
\reference{} Graham, A. W., Erwin, P., Caon, N., \& Trujillo, I. 2001, \apjl, 563, L11
%\reference{} Haiman, Z. \& Loeb, A. 2001, \apj, 552, 449
\reference{} Hamann, F. \& Ferland, G. 1992, \apjl, 391, L53
\reference{} Hamann, F. \& Ferland, G. 1993, \apj, 418, 11
%\reference{} Hamann, F. 1997, \apjs, 109, 279
%\reference{} Hamann, F. et al. 1997, \apj, 488, 155
\reference{} Hamann, F., Shields, J. C., Burbridge, E. M., Junkkarinen, V., \& Crenshaw, D. M. 1998, \apj, 496, 761
\reference{} Hamann, F. \& Ferland, G. 1999, \araa, 37, 487
\reference{} Hamann, F., Korista, K. T., Ferland, G. J., Warner, C., \& Baldwin, J. 2002, \apj, 564, 592 
\reference{} Hamann, F., Dietrich, M., Sabra, B., \& Warner, C. 2003, astro-ph/0306068 
\reference{} Henry, R. B. C., Edmunds, M. G., \& K\"{o}ppen, J. 2000, \apj, 541, 660
%\reference{} Jablonka, P., Martin, P., \& Arimoto, N. 1996, \aj, 112, 1415
\reference{} Kaspi, S., Smith, P. S., Netzer, H., Maoz, D., Jannuzi, B. T., \& Giveon, U. 2000, \apj, 533, 631
\reference{} Kellermann, K. I. et al. 1989, \aj, 98, 1195
\reference{} Korista, K.T. et al. 1995, \apjs, 97, 285
%\reference{} Kriss, G. 1994, in ASP Conf. Ser. 61, Astronomical Data Analysis Software and Systems III, eds. D. R. Crabtree, R. J. Hanisch, \& J. Barnes (San Francisco: ASP), 437
\reference{} Krolik, J. 2001, \apj, 551, 72
\reference{} Kuraszkiewicz, J., Wilkes, B., Czerny, B., \& Mathur, S. 2000, \apj, 542, 692
%\reference{} Laor, A. et al. 1994, \apj, 420, 110
\reference{} Laor, A., Jannuzi, B. T., Green, R. F., \& Boroson, T. A. 1997, \apj, 489, 656
\reference{} Laor, A. 1998, \apjl, 505, L83 
\reference{} Magorrian, J. et al. 1998, \aj, 115, 2285
\reference{} Mathur, S. 2000, \mnras, 314, 17
%\reference{} Marziani, P., Sulentic, J. W., Dultzin-Hacyan, D., Calvani, M., \& Moles, M. 1996, \apjs, 104, 37
\reference{} McLure, R. J. \& Dunlop, J. S. 2001, \mnras, 327, 199
\reference{} McLure, R. J. \& Dunlop, J. S. 2002, \mnras, 331, 795
\reference{} McLure, R. J. \& Jarvis, M. J. 2002, \mnras, 337, 109
\reference{} Merritt, D. \& Ferrarese, L. 2001, \apj, 547, 140
\reference{} Nelson, C. H. 2000, \apjl, 544, L91
\reference{} Netzer, H. \& Laor, A. 1993, \apjl, 404, L51
\reference{} Netzer, H. 2003, \apjl, 583, L5
%\reference{} Nussbaumer, H. \& Storey, P. J. 1982, \aap, 115, 205
%\reference{} Osmer, P., Porter, A. C., \& Green, R. F. 1994, \apj, 436, 678
%\reference{} Osmer, P. \& Shields, J. C. 1999, La Serena Conference on 'Quasars and Cosmology', eds. G. Ferland \& J. Baldwin
\reference{} Osterbrock, D. E. 1989, Astrophysics of Gaseous Nebulae and Active
Galactic Nuclei (Sausalito, California: University Science Books)
\reference{} Osterbrock, D. E. \& Pogge, R. W. 1985, \apj, 297, 166
\reference{} Peterson, B. M. 1993, \pasp, 105, 247
\reference{} Peterson, B. M. 1997, An Introduction to Active Galactic Nuclei (Cambridge: Cambridge University Press)
\reference{} Peterson, B. M. \& Wandel, A. 1999, \apjl, 521, L95
\reference{} Peterson, B. M. \& Wandel, A. 2000, \apjl, 540, L13
%\reference{} Petitjean, P., Rauch, M., \& Carswell, R. F. 1994, \aap, 291, 29
%\reference{} Pettini, M., Ellison, S. L., Bergeron, J., \& Petitjean, P. 2002, \aap, 391, 21
%\reference{} Pilyugin, L. S., Moll\'{a}, M., Ferrini, F., \& V\'{i}lchez, J. M. 2002, \aap, 383, 14
\reference{} Pilyugin, L. S. 2003, \aap, 399, 1003
\reference{} Pilyugin, L. S., Thuan, T. X., \& Vilchez, J. M. 2003, \aap, 397, 487
\reference{} Rees, M. J. 1984, \araa, 22, 471
\reference{} Shemmer, O. \& Netzer, H. 2002, \apjl, 567, L19
\reference{} Shemmer, O. et al. 2003, in ASP Conf. Ser. in press, AGN Physics
with the Sloan Digital Sky Survey, ed. G. T. Richards \& P. B. Hall
\reference{} Shields, G. A. 1976, \apj, 204, 330
\reference{} Shields, G. A. et al. 2003, \apj, 583, 124
\reference{} Steidel, C. C. et al. 2002, \apj, 576, 653
\reference{} Stern, D. et al. 2000, \aj, 119, 1526
\reference{} Stirpe, G. M. et al. 1994, \apj, 425, 609
%\reference{} Telfer, R. C., Zheng, W., Kriss, G. A., \& Davidsen, A. F. 2002, \apj, 565, 773
%\reference{} Trager, S. C., Faber, S. M., Worthey, G., \& Gonzalez, J. J. 2000, \aj, 120, 165
\reference{} Tremaine, S. et al. 2002, \apj, 574, 740
\reference{} Tinsley, B. 1980, Fundam. Cosmic Physics, 5, 287
%\reference{} van Albada, T. S., Bertin, G., \& Stiavelli, M. 1995, \mnras, 276, 1255
%\reference{} van Zee, L., Salzer, J. J., \& Haynes, M. P. 1998, \apjl, 497, L1 
\reference{} Vanden Berk, D. E. et al. 2001, \aj, 122, 549
%\reference{} Verner, E. M. et al. 1999, \apjs, 120, 101
\reference{} V\'{e}ron-Cetty, M.-P. \& V\'{e}ron, P. 2001, ``A Catalogue of Quasars and Active Nuclei", 10th edition
\reference{} Vestergaard, M. \& Wilkes, B. J. 2001,\apjs, 134, 1
\reference{} Vestergaard, M. 2002, \apj, 571, 733
\reference{} Vestergaard, M. 2004, \apj, in press 
\reference{} Wandel, A. 1999, \apjl, 519, L39
\reference{} Wandel, A., Peterson, B. M., \& Malkan, M. A. 1999, \apj, 526, 579
\reference{} Wang, T. \& Lu, Y. 2001, \aap, 377, 52
\reference{} Wang, J-M. 2003, \aj, 125, 2859
\reference{} Warner, C., Hamann, F., Shields, J. C., Constantin, A., Foltz, C., \& Chaffee, F. 2002, \apj, 567, 68 
\reference{} Warner, C., Hamann, F., \& Dietrich, M. 2003, \apj, 596, 72 
\reference{} Wheeler, J. C., Sneden, C., \& Truran, J. W. 1989, \araa, 27, 279
\reference{} Wills, B. J., Brotherton, M. S., Fang, D., Steidel, C. C., \& Sargent, W. L. W. 1993, \apj, 415, 563
\reference{} Wills, B. J. et al. 1995, \apj, 447, 139
\reference{} Wilkes, B. J. et al. 1999, \apj, 513, 76
%\reference{} Zaritsky, D., Kennicutt, R. C., \& Huchra, J. P. 1994, \apj, 440, 606
%\reference{} Zheng, W. \& Malkan, M. A. 1993, \apj, 415, 517
\reference{} Zheng, W., Kriss, G. A., Telfer, R. C., Grimes, J. P., \& Davidsen, A. F. 1997, \apj, 475, 469 
\end{references}
\end{document}